%% file: ms_2c.tex
\documentclass[12pt,preprint]{emulateapj}
\usepackage{natbib}
\input{colordvi}

\newcommand{\mic}{$\mu$m}
\newcommand{\spitzer}{{\it Spitzer}}

\slugcomment{Last edited: \today}

\shortauthors{P\'erez-Gonz\'alez et al.}

\begin{document}

\title{ULTRAVIOLET THROUGH FAR-INFRARED SPATIALLY RESOLVED ANALYSIS OF THE RECENT STAR FORMATION IN M81 (NGC3031)}

\author{Pablo G. P\'erez-Gonz\'alez\altaffilmark{1}, Robert C. Kennicutt Jr.\altaffilmark{1}, 
Karl D. Gordon\altaffilmark{1}, Karl A. Misselt\altaffilmark{1},
Armando Gil de Paz\altaffilmark{2,3}, Charles
W. Engelbracht\altaffilmark{1}, George H. Rieke\altaffilmark{1},
George J. Bendo\altaffilmark{4,1}, Luciana Bianchi \altaffilmark{5},
Samuel Boissier\altaffilmark{6,3}, Daniela Calzetti\altaffilmark{7},
Daniel A. Dale\altaffilmark{8}, Bruce T. Draine\altaffilmark{9},
Thomas H. Jarrett\altaffilmark{10}, David Hollenbach\altaffilmark{11},
Moire K. M. Prescott\altaffilmark{1}}

\altaffiltext{1}{The University of Arizona, Steward Observatory, 933 N Cherry Avenue, Tucson, AZ 85721}
\altaffiltext{2}{Departamento de Astrof\'{\i}sica, Facultad de CC. F\'{\i}sicas, 
Universidad Complutense de Madrid, E-28040 Madrid, Spain}
\altaffiltext{3}{Observatories of the Carnegie Institution of Washington, 813 Santa Barbara Street, Pasadena, CA 91101}
\altaffiltext{4}{Blackett Laboratory, Imperial College, London SW7 2AZ, UK}
\altaffiltext{5}{Department of Physics and Astronomy, Johns Hopkins University, 3400 North Charles Street, Baltimore, MD 21218}
\altaffiltext{6}{Laboratoire d'Astrophysique de Marseille, Traverse du Siphon-Les trois Lucs, BP8-13376 Marseille Cedex 12, France}
\altaffiltext{7}{Space Telescope Science Institute, 3700 San Martin Drive, Baltimore, MD 21218}
\altaffiltext{8}{Department of Physics and Astronomy, University of Wyoming, Laramie, WY 82071}
\altaffiltext{9}{Princeton University Observatory, Peyton Hall, Princeton, NJ 08544}
\altaffiltext{10}{Spitzer Science Center, California Institute of Technology, 100-22 IPAC, Pasadena, CA 91125}
\altaffiltext{11}{NASA Ames Research Center, Moffett Field, CA 94035}

\begin{abstract} 
The recent star formation (SF) in the early-type spiral galaxy M81 is
characterized using imaging observations from the far-ultraviolet (UV)
to the far-infrared (IR). We compare these data with models of the
stellar, gas, and dust emission for sub-galactic regions. Our results
suggest the existence of a diffuse dust emission not directly linked
to the recent SF. We find a radial decrease of the dust temperature
and dust mass density, and in the attenuation of the stellar
light. The IR emission in M81 can be modeled with three components: 1)
cold dust with a temperature $<T_c>=18\pm2$~K, concentrated near the
HII regions but also presenting a diffuse distribution; 2) warm dust
with $<T_w>=53\pm7$~K, directly linked with the HII regions; and 3)
aromatic molecules, with diffuse morphology peaking around the HII
regions. We derive several relationships to obtain total IR
luminosities from IR monochromatic fluxes, and we compare five
different star formation rate (SFR) estimators for HII regions in M81
and M51: the UV, H$\alpha$, and three estimators based on \spitzer\,
data. We find that the H$\alpha$ luminosity absorbed by dust
correlates tightly with the 24~\mic\, emission. The correlation with
the total IR luminosity is not as good. Important variations from
galaxy to galaxy are found when estimating the total SFR with the
24~\mic\, or the total IR emission alone. The most reliable
estimations of the total SFRs are obtained by combining the H$\alpha$
emission (or the UV) and an IR luminosity (especially the 24~\mic\,
emission), which probe the unobscured and obscured SF,
respectively. For the entire M81 galaxy, about 50\% of the total SF is
obscured by dust. The percentage of obscured SF ranges from 60\% in
the inner regions of the galaxy to 30\% in the outer zones.
\end{abstract}
\keywords{dust, extinction --- galaxies: individual (M81, NGC3031) --- galaxies: photometry --- galaxies: ISM --- galaxies: spiral --- galaxies: stellar content --- infrared: galaxies}

\section{INTRODUCTION}
\label{intro}

The multiwavelength analysis of the stellar populations in nearby
galaxies is one of the most powerful ways of understanding how
galaxies form and evolve \citep[see,
e.g.,][]{1999ApJ...521...64M,2000AJ....119...79C,2001PASP..113.1449C,
2003ApJ...587L..27P,2003ApJ...591..827P,2004MNRAS.351.1151B}. Indeed,
relatively bright and close objects are the best laboratories to study
galaxy evolution, given that we can obtain high quality data over a
wide range of wavelengths with incomparable depth and physical
resolution .The detailed characterization of nearby galaxies is
essential to establish the benchmark to which analysis of galaxies at
intermediate and high redshift must refer.

The multiwavelength data can be used to relate the scale where star
formation normally occurs (i.e., the typical size of an HII region,
20-200~pc; see, e.g., \citealt{1999ApJ...519...55Y}) to the scales of
the dynamical structures (bulge, arms, bars) and the entire galaxy
(i.e., several kpc). Moreover, the combination of data from the
ultraviolet (UV) to the far-infrared (FIR) and radio also allows
simultaneous analysis of the three main (emitting) components of a
galaxy: the stars, the dust, and the gas.

One of the main problems in the study of the stellar populations in
unresolved galaxies is the determination of the dust attenuation of
the stellar emission. Indeed, the correction for dust extinction is
the main source of uncertainty in the estimation of important
characteristics of entire galaxies (i.e., of integrated stellar
populations), such as the star formation rate (SFR), age or
metallicity \citep{1998ARA&A..36..189K,2000ApJ...533..236G,
2001PASP..113.1449C,2002ApJ...577..150B}. This problem is even more
severe when studying galaxies with current star formation, given that
this process occurs in zones where the gas and dust densities are high
(and the higher the densities, the more intense the star formation),
resulting in high extinctions with large spatial variations. This
effect translates into significant uncertainties concerning the
evolution of galaxies throughout the Hubble time
\citep[see][and references therein]{2004ApJ...615..209H}.

One way to overcome the problem of dust attenuation is to combine data
covering the widest wavelength range possible. This range must include
the UV, where the emission comes mainly from hot young massive stars,
and the optical and near-infrared, which are normally dominated by
more evolved low mass stars. Spectroscopic observations and
narrow-band imaging concentrating on emission-lines are also useful to
study the stellar populations (for example, the H$\alpha$ emission is
directly linked to the youngest stars). Much work characterizing the
stellar populations in galaxies using a variety of such datasets and
technical approaches has appeared in the literature
\citep[among others,][]{2001ApJ...550..212B,
2001ApJ...559..620P,2003MNRAS.338..525P,2003MNRAS.341...33K,
2004Natur.428..625H}.

A complementary way to study the star formation (especially the most
recent) in galaxies is using mid- and far-infrared observations. The
emission at these wavelengths comes mainly from dust heated by stellar
light \citep[see, e.g.,][]{1996ApJ...460..696W,1998ARA&A..36..189K,
1999A&A...350..381D}. The total IR emission (integrated from 8 to
1000~\mic) has been claimed to be an optimum SFR tracer for entire
galaxies (at least for the ones dominated by star formation and with
high dust content), given that it is not affected by dust attenuation
\citep[see][and references therein]{1992ApJ...396L..69S,
1996ARA&A..34..749S,1998ARA&A..36..189K}. The IR has been widely used
for studying local and distant star-forming galaxies
\citep[e.g.,][]{1972ApJ...176L..95R,1990MNRAS.242..318S,1999ApJ...517..148F,
1999ApJ...521...64M,2001A&A...378....1F,2001ApJ...549..745R,
2001ApJ...556..562C,2002AJ....124.3135K,2002ApJ...568..651G,
2002ApJ...576..159D,2005ApJ...630...82P}. Although it is generally
assumed that the heating of the dust emitting in the IR is dominated
by light coming from the young hot stars (which would mean that the IR
is a robust SFR estimator), it remains unclear how much the older
stellar populations contribute (i.e., how we should correct the
integrated IR emission to obtain robust values of the current
SFR). Moreover, the validity of the equations correlating the SFR and
the IR emission, which were derived for entire galaxies, is not well
established when working at sub-galactic scales (for example, with HII
regions). Finally, another interesting topic under study is the role
of the aromatic molecules (loosely termed polycyclic aromatic
hydrocarbons, PAHs) in the star formation event (see
\citealt{1984A&A...137L...5L}, \citealt{1985ApJ...290L..25A},
\citealt{1989ARA&A..27..161P}; and most recently, 
\citealt{2004A&A...419..501F,2004A&A...428..409B,2004ApJ...613..986P}).

The study of the stellar populations in galaxies, and more
specifically of the most recent star formation, has experienced
remarkable progress in the past few years, as we have been able to
gather observations from the UV to the FIR with similar depths and
spatial resolutions. Two space missions have contributed significantly
to this observational advance: the Galaxy Evolution Explorer (GALEX;
\citealt{2005ApJ...619L...1M}) and the
\spitzer\, Space Telescope \citep{2004ApJS..154....1W}, which have
unprecedented sensitivity and angular resolution and coverage in the
UV and IR, respectively. These facilities, jointly with other
ground-based and space telescopes, are being used by the \spitzer\,
Infrared Nearby Galaxy Survey (SINGS, \citealt{2003PASP..115..928K})
to collect very high quality data from the UV to the FIR for a large
sample of nearby resolved galaxies representative of the local galaxy
population. The main goal of SINGS is to provide a complete
multiwavelength dataset to understand the formation of stars in
galaxies and the properties of the dust. The SINGS dataset, including
up to 7 broad-band measurements in the IR, and extensive ancillary
data at optical and radio wavelengths, has proved to be very useful
for exploring these topics
\citep[see, e.g.,][]{2005ApJ...633..857D,2005ApJ...633..871C}.

In this paper, our main goal is to study the characteristics of the
most recent star formation in a galaxy, testing the IR emission as a
SFR estimator for sub-galactic regions by empirically analyzing its
correlation with the ``classical'' star formation tracers, the UV
continuum and the ionized hydrogen emission-lines (in our case, the
H$\alpha$ emission). This paper complements a similar analysis of M51
by \citet{2005ApJ...633..871C}. Combining the data for the three SFR
estimators, we can account for: 1) the photons coming from the
newly-formed stars which do not interact with anything or are just
scattered, and we detect in the UV; 2) the photons that interact with
the gas, and are reemitted in emission lines; and 3) the photons that
interact with the dust, and are reemitted in the IR. By checking the
consistency among these indicators, a clearer picture of the star
formation event is obtained. The results in this paper will be used to
study the different stellar populations (i.e., the star formation
history) of M81 in a future work (P\'erez-Gonz\'alez et al. 2006, in
preparation).

This work concentrates on one of the galaxies in the SINGS sample, M81
(NGC3031), the largest in angular extent in the survey (with a size of
$27\arcmin\times14\arcmin$). M81 is a face-on spiral [SA(s)ab type,
\citealt{1992yCat.7137....0D}] at a distance of 3.63$\pm$0.34~Mpc 
\citep{1994ApJ...427..628F,2002A&A...383..125K}. At this
distance, the resolution of the MIPS images corresponds to 100-700~pc
(depending on the channel). This resolution allows the study of the
star formation in M81 at different scales ranging from typical sizes
of HII regions to the size of large star-forming complexes and
dynamical structures (bulge, arms). M81 is an object with moderate
global star formation
($SFR=0.4-0.8$~$\mathcal{M}_\sun\,\mathrm{yr}^{-1}$;
\citealt{1995AJ....110.1115D}, \citealt{2003AJ....126.1286L}, 
\citealt{2004ApJS..154..215G}) and extinction ($A(V)=0.5-1.0$, 
\citealt{1987ApJ...317...82G}, \citealt{2000AJ....119.2745K}). It 
forms part of a group of about 25 galaxies with clear interactions
among them (for example, among M81, M82, and NGC3077; see
\citealt{1977ApJ...211..707S}). M81 also harbors an active galactic
nucleus \citep{1981ApJ...245..845P, 1988ApJ...324..134F} classified as
a LINER Sy1.5-1.8 \citep{1980A&A....87..152H,1981ApJ...250...55S,
1997ApJS..112..315H}. The nucleus is a strong x-ray
\citep{1982ApJ...257L..51E, 1988ApJ...325..544F,
1993ApJ...418..644P,2003A&A...400..145P, 2004ApJ...601..831L} and UV
\citep{1980ApJ...237..290W, 1992ApJ...395L..37H,1994A&AS..106..523R}
emitter. M81 has been extensively studied at all wavelengths from
x-rays to the radio. In this work, we will make use of the UV and
emission-line data published for some of the individual HII regions in
M81 \citep{1987ApJ...317...82G,1987ApJ...319...61K,
1995ApJ...438..181H,2003AJ....126.1286L}, jointly with the new {\it
Spitzer} and GALEX data.

This paper is organized as follows. Section~\ref{data} describes the
data compiled for this work. In Section~\ref{modeling}, we present the
models of the stellar and dust thermal emission developed to analyze
the UV-to-FIR data of M81. In Section~\ref{results_1}, we discuss the
contribution of stellar light to the mid-IR fluxes, and different ways
to estimate the total IR luminosity from monochromatic fluxes measured
by {\it Spitzer}. We also concentrate on the analysis of the dust
properties in several sub-galactic regions in
M81. Section~\ref{results_2} compares different SFR estimators for
selected HII regions. Section~\ref{conclusions} summarizes our
conclusions.

\section{THE DATA}
\label{data}

We have compiled a variety of imaging and spectroscopic data for M81
covering the ultraviolet (UV), optical, near-IR (NIR), mid-IR (MIR),
and far-IR (FIR) wavelengths. The sources for each piece of the
dataset are described in the following subsections.

\subsection{The \spitzer\, data}

Observations of M81 in the MIR and FIR were carried out with
\spitzer\, \citep{2004ApJS..154....1W} as a part of the SINGS 
Legacy Project \citep{2003PASP..115..928K}. The Infrared Array Camera
(IRAC, \citealt{2004ApJS..154...10F}) obtained images of M81 at 3.6,
4.5, 5.8, and 8.0~\mic\, in May 2004. The data were processed with
version S10.0.3 of the Spitzer Science Center pipeline, and then
mosaicked to produce a single image for each channel with a scale of
0.76~arcsec~pixel$^{-1}$ \citep{2004ApJS..154..204R}. The Full Width
at Half Maximum (FWHM) of the Point Spread Function (PSF) is
1.7$\arcsec$, 1.7$\arcsec$, 1.9$\arcsec$, and 2.0$\arcsec$ for the
3.6, 4.5, 5.8, and 8.0~\mic\, channels, respectively (corresponding to
a physical size of $\sim$30-35~pc at the distance of M81). The IRAC
images were calibrated using the factors and aperture corrections (to
properly measure the photometry of extended sources) found in
\citet{2005PASP..117..978R}. These corrections range from 6\% in 
channel 1 to 26\% in channel 4. The 1-$\sigma$ sensitivities of the
images are $[0.2,2.2,2.0,1.9]$~$\mu$Jy~arcsec$^{-2}$ at
$[3.6,4.5,5.8,8.0]$~\mic\, ($[25.7,23.0,23.2,23.2]$~mag(AB)~arcsec$^{-2}$).

Images at 24, 70 and 160~\mic\, were taken with the Multiband Imaging
Photometer for \spitzer\, (MIPS,
\citealt{2004ApJS..154...25R}) at three different
epochs (to remove artifacts) in November 2003 and October 2004. The
three epochs were processed and mosaicked together using the DAT
reduction pipeline \citep{2005PASP..117..503G}, producing final
mosaics with pixel scales half of the original values (1.25, 4.9, and
8~arcsec~pixel$^{-1}$ for 24, 70, and 160~\mic, respectively). These
mosaics have a size of $1\arcdeg\times2.5\arcdeg$, large enough to
include the entire galaxy.  The FWHM of the PSFs are 5.7$\arcsec$,
16$\arcsec$, 38$\arcsec$ for the 24, 70, and 160~\mic\, channels
(corresponding to 100, 282, and 669~pc in M81). The 1-$\sigma$
sensitivities of the images are $[0.9,7.0,20.0]$~$\mu$Jy~arcsec$^{-2}$
at $[24,70,160]$~\mic\, ($[24.0,21.8,20.7]$~mag(AB)~arcsec$^{-2}$).

\subsection{The GALEX data}

The Galaxy Evolution Explorer (GALEX) observed M81 on December
8$^{th}$ 2003 for a total of 3089\,seconds split across two
orbits. Note that since GALEX only observes during night-time the
usable time per orbit is only $\sim$1700\,seconds. Images in the two
GALEX bands, far-ultraviolet (FUV; $\lambda_{\mathrm{eff}}$=151.6~nm)
and near-ultraviolet (NUV; $\lambda_{\mathrm{eff}}$=226.7~nm), were
obtained simultaneously. The UV light is separated in these two bands
by a dichroic beam splitter that also acts as a field-aberration
corrector. Individual UV photons are then detected using micro-channel
plates with crossed delay-line anodes. The photon lists created are
processed by the GALEX pipeline to produce the corresponding intensity
maps in counts per second with a final scale of 1.5
arcsec\,pixel$^{-1}$ (see \citealt{2005ApJ...619L...7M}). The reader
is referred to \citet{2005ApJ...619L...1M} for a more detailed
description of the GALEX mission.

The large field of view of the GALEX instrument (a circle of
1.2$\arcdeg$ in diameter) allowed it to cover the entire M81 galaxy,
along with M82 and Holmberg~IX, in one single pointing. The absolute
flux calibration of the GALEX data is based on multiple observations
of white dwarf standard stars. Zero-point errors are estimated to be
of the order of 0.15\,mag both in the FUV and NUV channels. The
astrometry of the GALEX images is based on the Tycho-2 catalog
\citep{2000A&A...357..367H} and it is found to have
a RMS error less than 1\,pixel. The FWHM of the PSF of the GALEX
images is a function of the brightness of the source and the position
on the detector. In the case of the observations of M81 the FWHM is
approximately 5$\arcsec$ and 6$\arcsec$ (88 and 106~pc at the M81
distance, respectively), respectively for the FUV and NUV bands. The
1-$\sigma$ sensitivities of the images of M81 are
$[1.8,1.7]\times10^{-2}$~$\mu$Jy~arcsec$^{-2}$
($[28.3,28.3]$~mag(AB)~arcsec$^{-2}$) for the FUV and NUV channels,
respectively.

\subsection{Optical and near-infrared data}

We complement the UV GALEX data and the MIR/FIR \spitzer\, images with
optical and NIR data obtained at different ground-based
facilities. $UBVRI$ observations were carried out in 2004-2005 with
the 2.3m Bok Telescope at Steward Observatory using the prime focus
wide-field imager \citep{2004SPIE.5492..787W}. Standard reduction
algorithms for CCD images were applied to the data. We took special
care with the distortion correction of the large
1$\arcdeg\times$1$\arcdeg$ images. The depths of these optical images
are $[7.1,7.5,11.0,12.0,2.6]\times10^{-2}$~$\mu$Jy~arcsec$^{-2}$
($[26.8,26.7,26.3,26.2,27.9]$~mag(AB)~arcsec$^{-2}$) for the
$UBVRI$ bands (1-$\sigma$ sensitivity), respectively.

NIR images ($JHK_s$ bands) were taken from the Two Micron All Sky
Survey \citep[2MASS;][]{2000AJ....119.2498J}. The depths are
$[80,140,140]$~$\mu$Jy~arcsec$^{-2}$
($[19.1,18.5,18.5]$~mag(AB)~arcsec$^{-2}$) in $JHK$ (1-$\sigma$
sensitivity), respectively.

Finally, we have used the narrow-band H$\alpha$ image taken from
\citet{1998ApJ...506..135G}. The calibration of this image was carried 
out by measuring the H$\alpha$+$[NII]$ fluxes for all the HII regions
cataloged in \citet{1987ApJ...319...61K} and
\citet{2003AJ....126.1286L}. Our calibration gives a total 
H$\alpha$+[NII] flux for M81 of
$4.7\pm1.0$~erg\,s$^{-1}$\,cm$^{-2}$. The difference between this flux
and the estimations of
\citet[][$4.8\pm0.7$~erg\,s$^{-1}$\,cm$^{-2}$] {1998ApJ...506..135G}
and
\citet[][$4.53\pm1.35$~erg\,s$^{-1}$\,cm$^{-2}$]
{1995AJ....110.1115D} is less than 4\% and well within errors. In the
following discussion, we will use pure H$\alpha$ fluxes that have been
measured from the narrow-band image and corrected for $[NII]$
contamination assuming an average $[NII]$/H$\alpha$ ratio of 0.4
\citep{1987ApJ...317...82G}. The 1-$\sigma$ H$\alpha$ sensitivity
limit of this image is
$1\times10^{-17}$~erg~s$^{-1}$~cm$^{-2}$~arcsec$^{-2}$.

\subsection{Photometry}

\input{tab1_2c}
\input{tab2_2c}

\begin{figure*}
\begin{center}
\includegraphics[width=7.6cm]{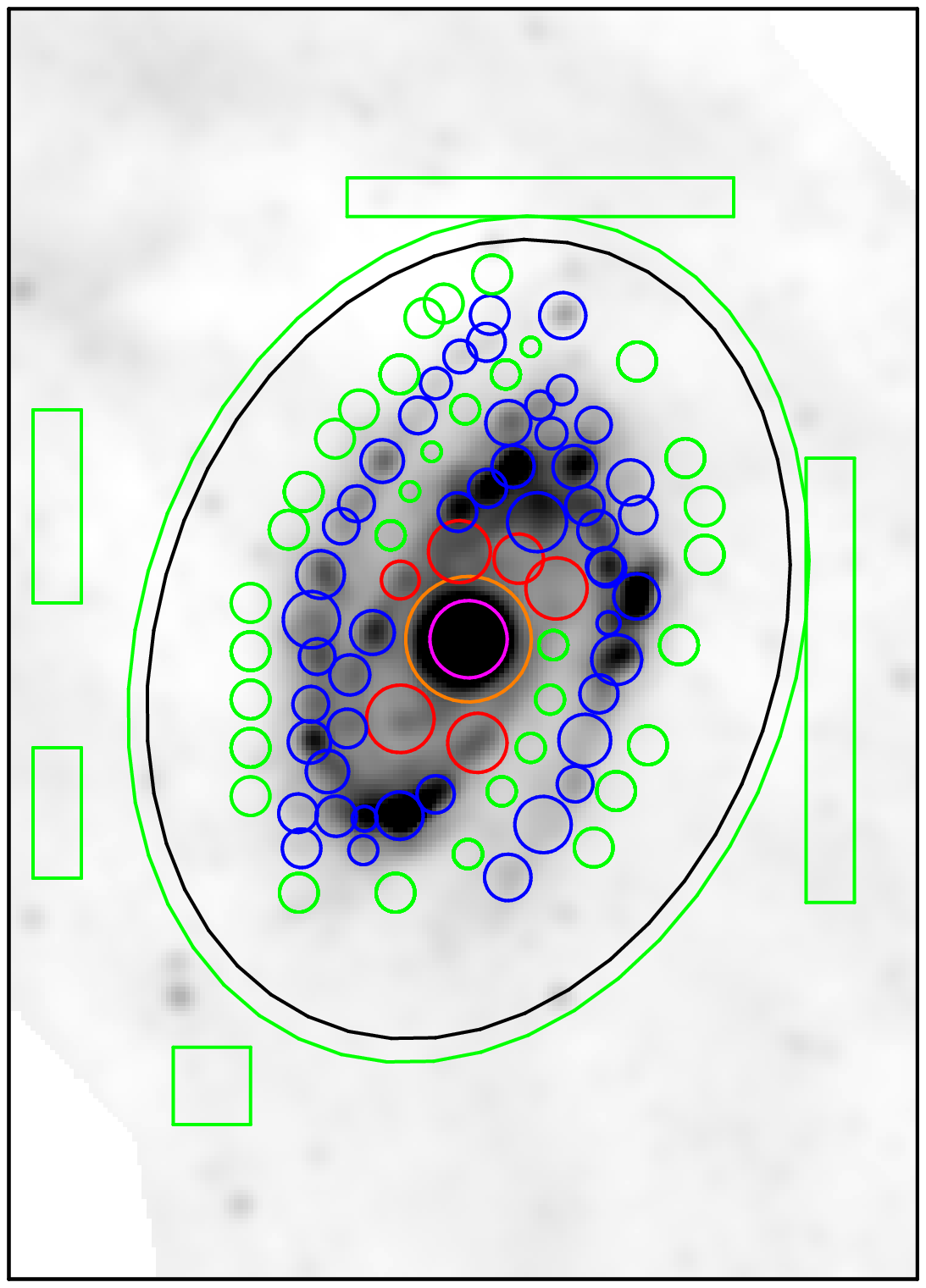}
\hspace{0.5cm}
\includegraphics[width=7.6cm]{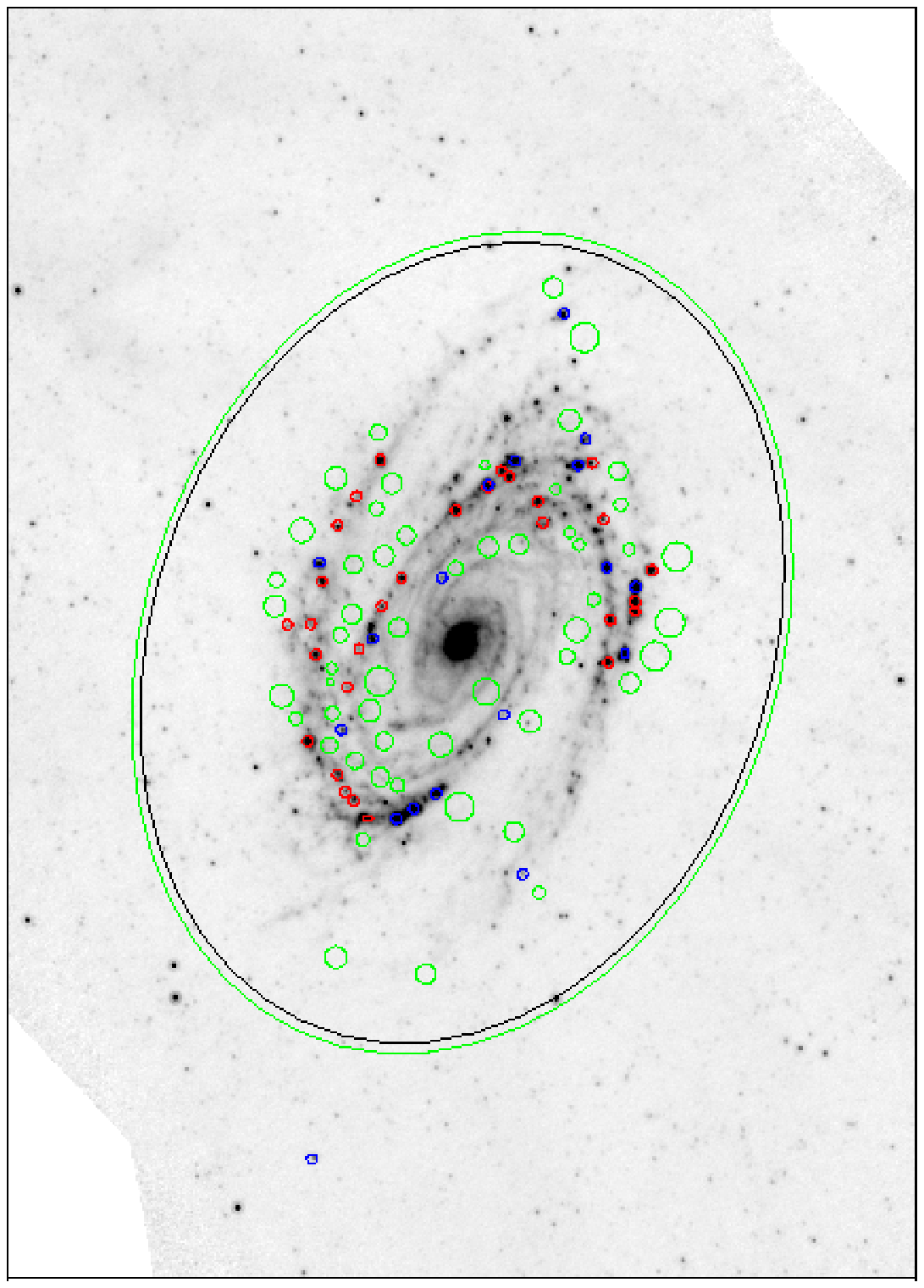}
\figcaption{\label{reg_160} The regions for which we extracted 
photometry from M81 are shown superimposed on the MIPS 24~\mic\,
image. The left panel shows the effect of convolving to the 160~\mic\,
resolution, and the right panel show the photometric apertures for the
HII regions selected at the original 24~\mic\, resolution. On the left
panel, blue circles enclose star-forming complexes in the arms. The
red circles enclose regions of diffuse emission in the zone between
the nucleus and the arms. The magenta region encloses the inner
nucleus of M81. The area outside the magenta circle and inside the
orange circles is defined as the circumnuclear region. Green
rectangles show the zones used for estimating the global background
(selected to be free of bright sources in all the images). For the
right panel, blue regions come from the sample of HII regions studied
by
\citet{1987ApJ...317...82G}, and red regions were found in
\citet{1987ApJ...319...61K}. For both panels, the black ellipse
depicts the aperture used for the entire M81 galaxy. The green ellipse
and circles show the zones where the local sky was measured (see text
for details).}
\end{center}
\end{figure*}

\slugcomment{Please, plot this figure with the width of one column}
\placefigure{reg_160}

We produced two sets of images, one set composed of images convolved
to match the PSF of the image in our dataset with the worst
resolution, the MIPS 160~\mic\, image, and another set convolved to
the MIPS 24~\mic\, PSF, which is adequate to study the dust emission
in individual HII regions. The World Coordinate System (WCS) for the
UV, optical, NIR, and IRAC images was determined using 2MASS stars
(more than 30 per frame). For the MIPS images, given that stars are
normally very faint or undetected, we determined the WCS by using
bright point-like regions detected in the IRAC bands. The images were
aligned and cropped to the same size using the WCS for each frame.

For the image set convolved to the 160~\mic\, resolution (160RES set
hereafter), we selected 49 circular regions in the arms corresponding
to resolved peaks in the 160~\mic\, emission. We also selected 6
regions in the zone between the arms and the nucleus dominated by a
more diffuse emission, two regions around the nucleus (one enclosing
the inner nucleus, and another the circumnuclear region), and an
aperture enclosing the entire galaxy. The radii of the photometric
apertures were chosen to enclose most of the emission from the peaks
(ranging from 0.6 to 1.8~kpc). The photometric apertures are shown in
the left panel of Figure~\ref{reg_160}. Most of the selected regions
were identified in the 24~\mic\, and H$\alpha$ images with bright
star-forming complexes, normally enclosing several individual HII
regions (see right panel of Figure~\ref{reg_160}).

Proper background subtraction is essential when dealing with data
spanning a wide range of wavelengths, especially if some of the images
include very bright diffuse emission (such as the images of M81 in the
MIR and FIR, see \citealt{2004ApJS..154..215G}). We followed two
approaches in the background estimation, each of them optimized for
one of the two problems that we can address with this dataset. First,
analyzing the properties of the global stellar population and the dust
in several sub-galactic regions in M81 requires photometry that
accounts for the entire emission of the region at any wavelength
(including all stars, gas and dust in the region), so the best
approach is to estimate a global background, i.e., a background that
only includes the sky emission around the galaxy. However, second, for
studying different SFR estimators, one of the main goals of this
paper, it is important to account for the diffuse emission surrounding
each star-forming complex. This emission is probably not directly
related to the recent star formation, but arises from a source in the
galaxy of a different origin (e.g., older stars, diffuse dust
emission). 

As an example, the 160~\mic\,image of M81 shows a bright diffuse
component that is not linked with any star-forming region (detected in
the H$\alpha$ or 24~\mic\, image). Other authors have also detected an
extended and diffuse dust component beyond the star-forming disk in a
variety of galaxies \citep[see, among
others,][]{2003A&A...407..137H,2003A&A...410L..21P,2004ApJS..154..259H,
2005A&A...443..373S,2005ApJ...619L..75P}. The cold dust responsible
for this emission is probably not (only) heated by the newly-formed
stars, but by a more diffuse heating source, such as an evolved
stellar population or Ly$\alpha$ and Lyman continuum photons escaping
from HII regions \citep[see][]{2005ApJ...619L..75P}.

The global background for M81 was measured in several zones located
around the galaxy, which were observed at all wavelengths and clear of
bright objects (green rectangles and elliptical annulus in the left
panel of Figure~\ref{reg_160}). The local background was measured in
small circular apertures around the arms (green circles in the left
panel of Figure~\ref{reg_160}). For each individual region studied, we
averaged the values obtained for the two closest sky apertures, one in
the inter-arm region and another in the outer part of the arm. Two
photometric catalogs were produced using the two background
estimations (marked as LOCALBKG and GLOBALBKG photometry), and the
implications of each choice are discussed in the following Sections.

For the image set convolved to the 24~\mic\, resolution (24RES
hereafter), we extracted photometry for 59 individual HII regions
found in
\citet{1987ApJ...317...82G} and \citet{1987ApJ...319...61K}. These regions 
are shown in the right panel of Figure~\ref{reg_160}. The original
aperture sizes used by those authors were very small compared with the
resolution of our 24~\mic\, image, so we extracted photometry for
circular apertures with a 7 pixels ($\sim$8.4$\arcsec$) radius, which
enclosed most of the emission for each HII region as seen in the
H$\alpha$ image (the radius of these apertures corresponds to
0.1~kpc), and is an adequate match to the 24~\mic\, PSF. The sky
estimation was performed locally in circular apertures near the
individual HII regions \footnote{Only one approach of the background
determination was followed for the 24RES case (local background),
given that we were interested in comparing the SFR estimators. This
task benefits from a proper subtraction of the diffuse emission
surrounding the HII regions.} (green circles in the right panel of
Figure~\ref{reg_160}). For the entire galaxy, the sky was measured in
an elliptical annulus.

Aperture corrections were applied to the fluxes of the regions
selected in each image set to obtain the photometry for an ``infinite
aperture'' using theoretical STinyTim PSFs for the MIPS bands.  All
the fluxes were corrected for Galactic extinction using a value of
$A(B)=0.346$
\citep{1998ApJ...500..525S} and the extinction law published by
\citet{1989ApJ...345..245C} with $R_V=3.1$. The fluxes for all 
regions are given in Tables~\ref{photometry_160RES} and
\ref{photometry_24RES}.



\section{MODELS OF THE UV-TO-FIR EMISSION OF M81}
\label{modeling}

\subsection{The stellar and gas emission model}

For the regions selected in the 160RES image set, we built spectral
energy distributions (SEDs) from 0.150~\mic\, to 160~\mic. SEDs were
also built for the HII regions selected in the 24RES image set from
0.150~\mic\, to 24~\mic. We then fitted the SEDs with models of the
stellar, gas, and dust emission. These models will be used in this
paper to remove the stellar contribution to the emission in the MIR
and FIR bands. In a future paper, the results on the stellar
populations of M81 obtained with these models will be discussed in
detail.

The photometric points at wavelengths bluer than 4.5~\mic\, were
fitted to stellar population synthesis models. We also included the
H$\alpha$ equivalent width in the fits
\citep[see][]{2000MNRAS.316..357G,2003MNRAS.338..508P,2003MNRAS.338..525P}.


We assumed that the star formation history in each region can be
described by a burst with an exponentially declining star formation
rate with time scale $\tau$, metallicity $Z$, age $t$, and attenuated
by an amount described by the quantity $A(V)$. The attenuation at any
wavelength was calculated from the free parameter $A(V)$ using the
\citet{2000ApJ...539..718C} recipe. The stellar emission in our models
was taken from the PEGASE code \citep{1997A&A...326..950F}, and we
added the emission from the gas heated by the stars (emission lines
and nebular continuum) using the emission and recombination
coefficients given by \citet{1980PASP...92..596F} for an electron
temperature $T_e=10^4\,K$, the relations given by
\citet{1971MNRAS.153..471B}, and the theoretical line-ratios expected
for a low density gas ($n_{\mathrm{e}}=10^{2}\,\mathrm cm^{-3}$) with
$T_{\mathrm{e}}=10^{4}\,\mathrm K$ in the recombination Case B
\citep{1989agna.book.....O}. We fixed the metallicity of our models to 
$Z_\sun$, based on the results from \citet{1994ApJ...420...87Z},
\citet{1984AJ.....89.1702S}, and \citet{2000AJ....119.2745K}, who
showed that the metallicity in M81 is nearly solar with a very weak
radial dependence. We probed the solution space in the following
ranges for the three remaining free parameters $[\tau,t,A(V)]$: 1) we
assumed $\tau$ values from an almost instantaneous burst
($\tau=1$~Myr) to an almost constant SFR ($\tau=15$~Gyr) using a
logarithmic interval of 0.1dex (in yr); 2) ages were probed from
$t=1$~Myr to $t=13$~Gyr in logarithm intervals; 3) extinction values
ranged from $A(V)=0$ to $A(V)=3$ in intervals of 0.05~mag.


\subsection{The model of emission by dust}

After fitting the UV-to-NIR part of the SEDs with stellar/gas emission
models, we subtracted the contribution from the stars/gas to the
fluxes in the MIR and FIR. We then fitted models of dust emission to
the remaining fluxes. These models are described by an equation of the
form:

\begin{equation}
\label{dustmodel}
F_\mathrm{dust}(\lambda) = \sum C^{i} \kappa^{i}(\lambda) B_{\lambda}(T_{\mathrm{dust}}^i)
\end{equation}

\noindent where the sum extends over the number of dust components 
(each of them emitting as a modified blackbody),
$C^{i}=M_{\mathrm{dust}}^i/D^{2}$, $D$ is the luminosity distance to
the source, $M_{\mathrm{dust}}^i$ are the masses of each dust
component, and $\kappa_i$ are the wavelength dependent mass absorption
coefficients. We adopted a model with 3 components: 1) warm silicates;
2) cold silicates (both components with grains of sizes
$a\sim0.1$~\mic); and 3) aromatic molecules. Mass absorption
coefficients for astronomical silicates were computed from Mie theory
using the dielectric functions of \citet{1993ApJ...402..441L}. Cross
sections for the aromatic molecules were taken from
\citet{2001ApJ...554..778L}. As the canonical aromatic spectrum \citep[see,
e.g.,][]{2004ApJS..154..309W} exhibits no features beyond 20~\mic, we
re-computed the aromatic cross-sections (and mass absorption
coefficients) leaving off the last three terms of Equation~11 in
\citet{2001ApJ...554..778L}. The aromatic component is included mainly for
completeness.  The aromatic emission is a stochastic process rather
than an equilibrium process as assumed in
Equation~\ref{dustmodel}. This means that our model should
underestimate the total aromatic masses. A proper treatment of the
stochastic aromatic emission would require a full radiative transfer
solution to derive the radiation field at each position in the galaxy.
The attendant assumptions are beyond the scope of this paper, but will
be addressed in forthcoming detailed models of individual regions. Our
results are not affected by this issue, given that they only refer to
the total dust masses, to which aromatic molecules make only a minor
contribution.

\begin{figure*}
\begin{center}
\includegraphics[width=7.cm,angle=-90]{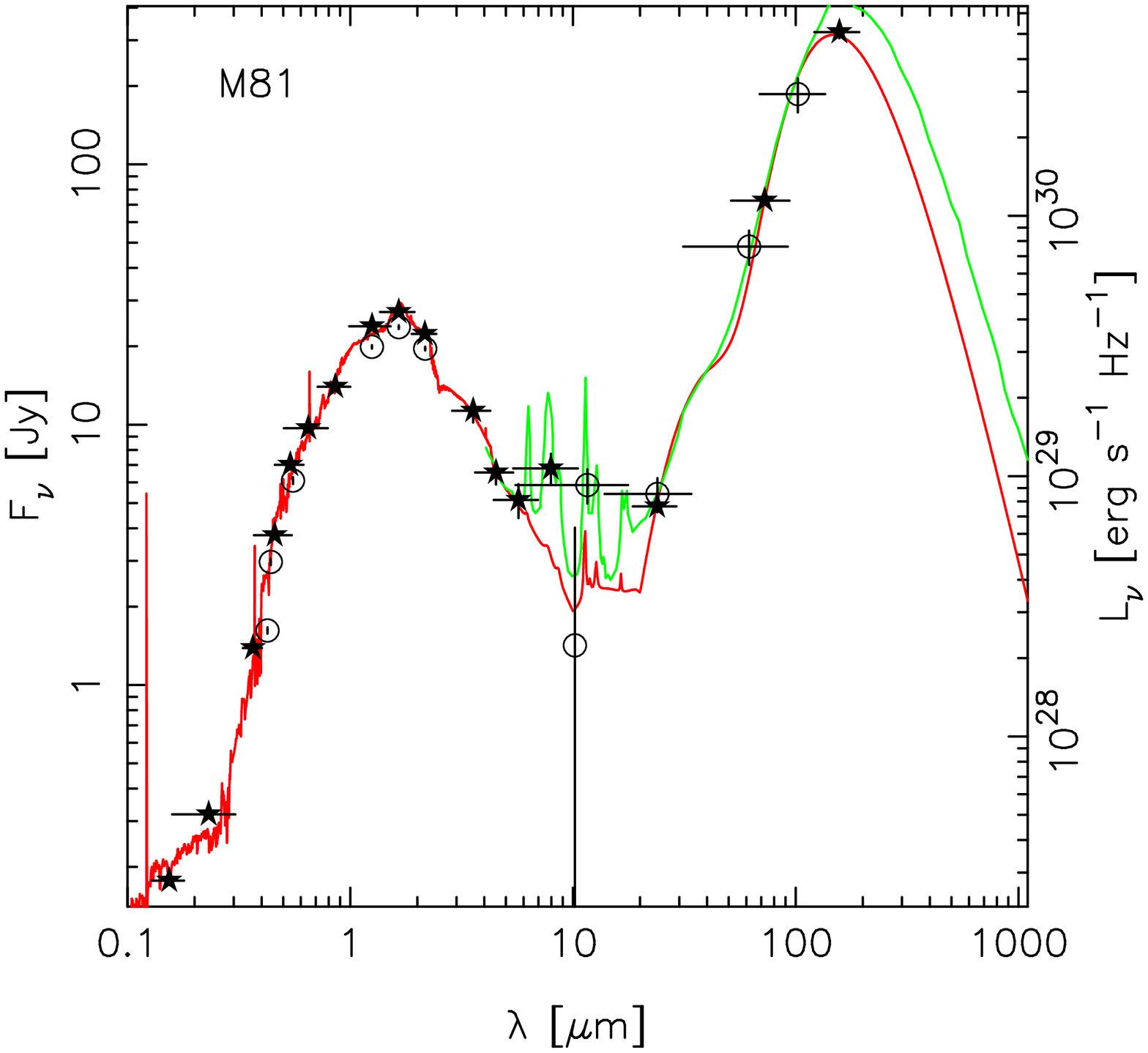}
\hspace{0.2cm}
\includegraphics[width=7.cm,angle=-90]{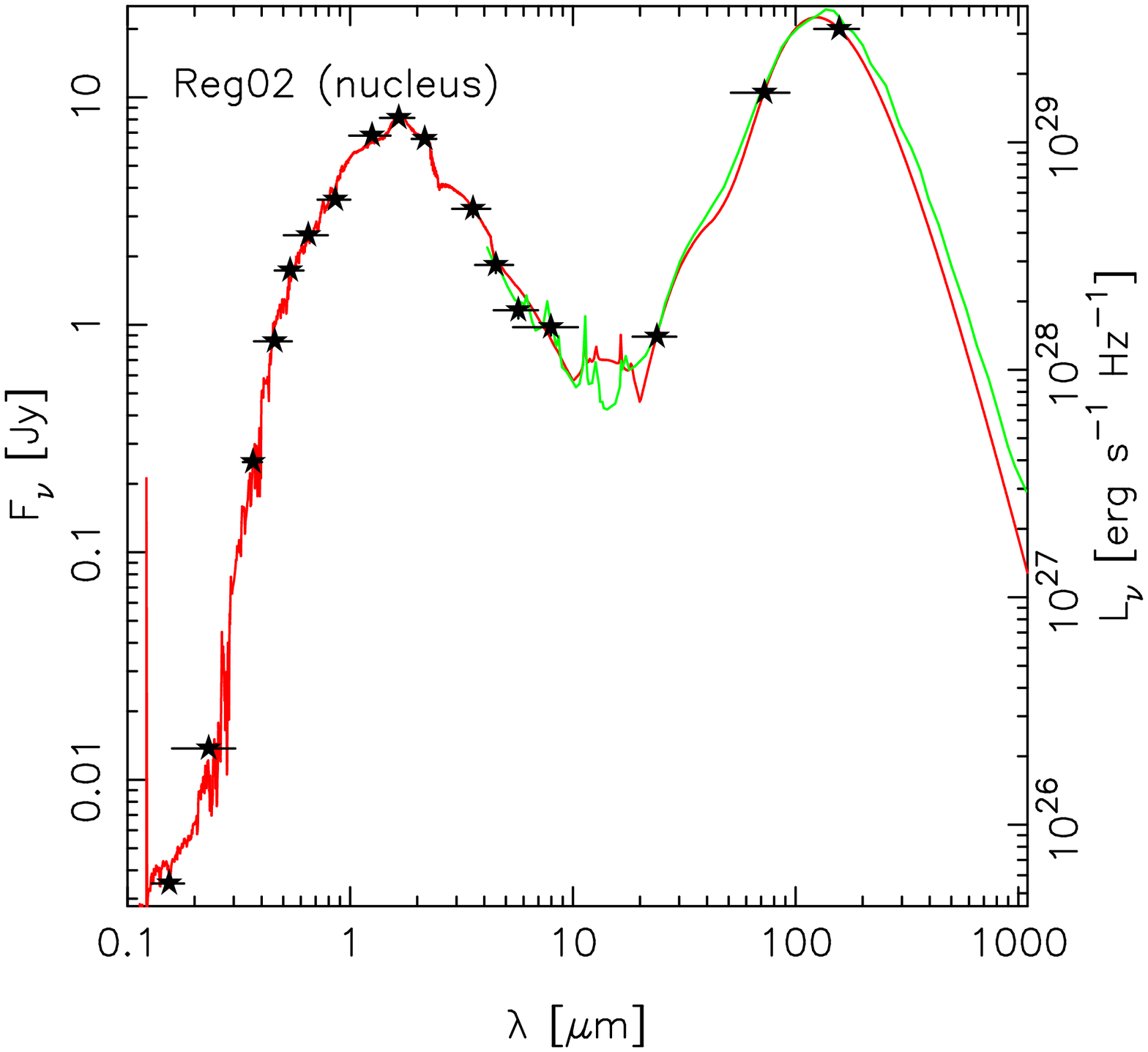}
\includegraphics[width=7.cm,angle=-90]{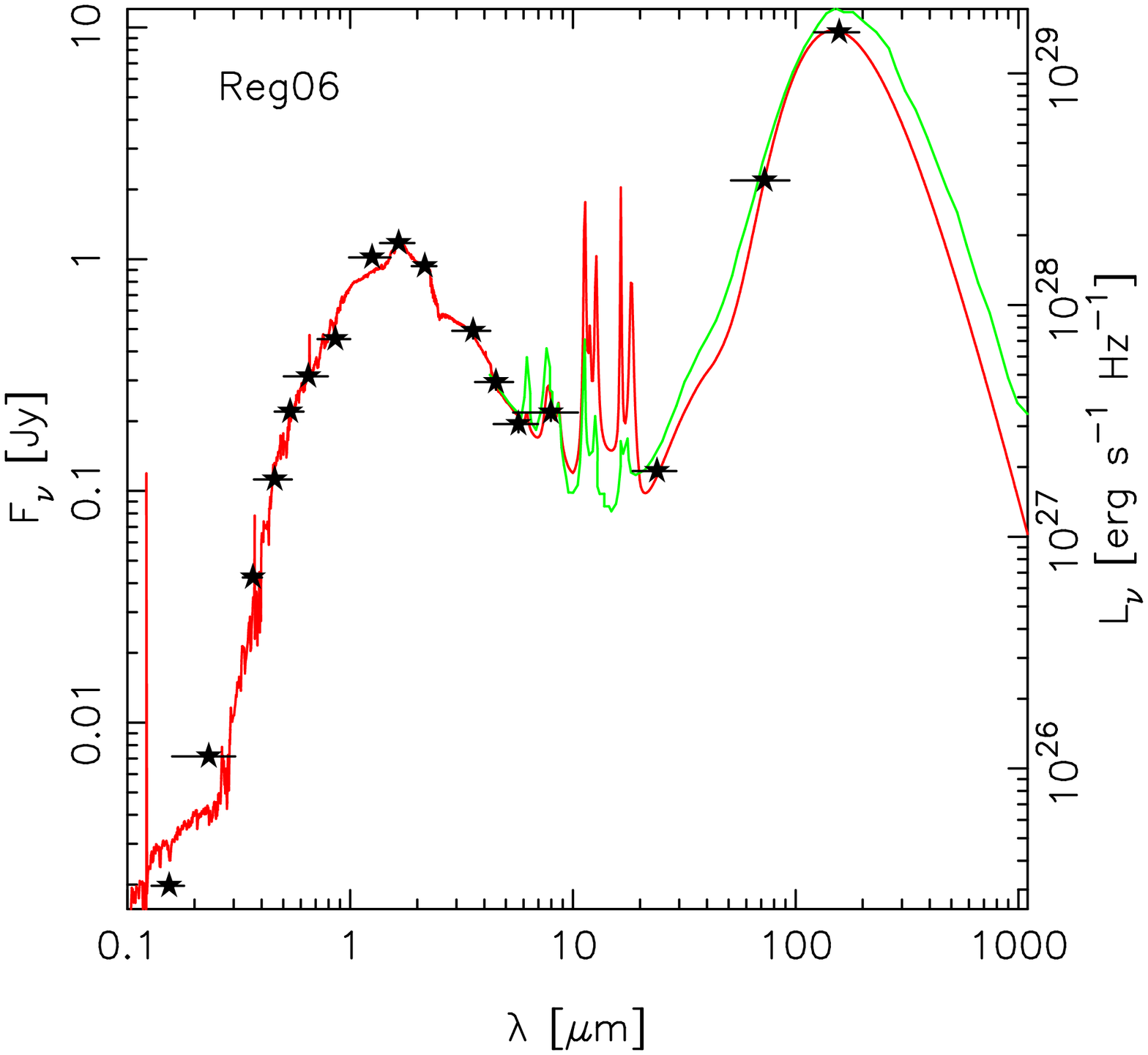}
\hspace{0.2cm}
\includegraphics[width=7.cm,angle=-90]{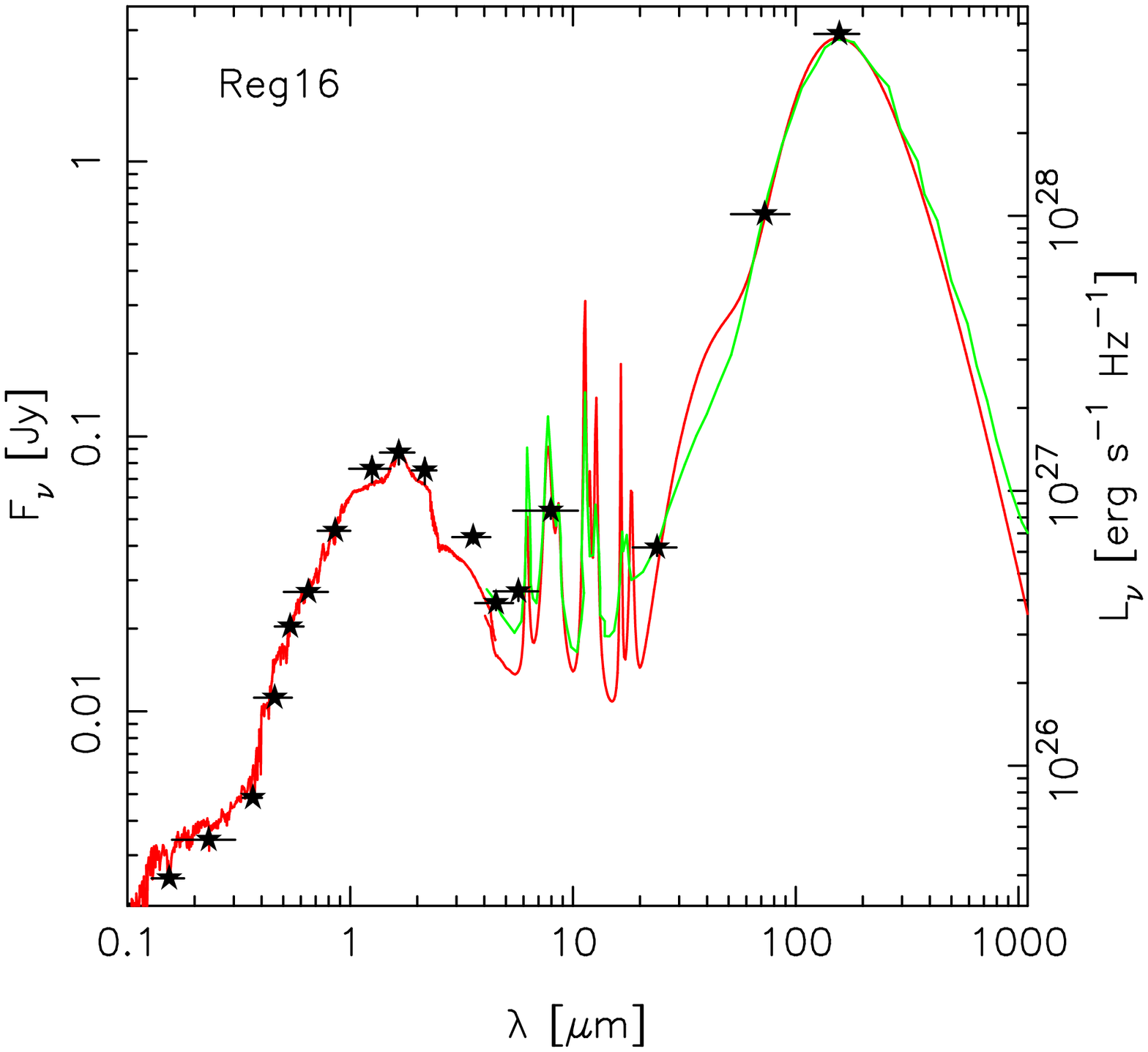}
\figcaption{\label{some_fits} Spectral energy distributions
for the entire M81 galaxy (upper-left panel), the nucleus
(upper-right), and two regions in the arms (lower panels) selected at
160RES, corresponding to the photometry estimated with the global sky
subtraction. The open circles in the SED of the entire galaxy refer to
several photometric points extracted from NED (for apertures nearly as
large as ours): three UV/optical fluxes from
\citet{1982ApJ...256....1C} and \citet{1992yCat.7137....0D}, the 
three NIR fluxes from 2MASS \citep{2000AJ....119.2498J}, the flux in
the $N$-band from \citet{1970ApJ...161L.203K}, and the four IRAS
fluxes measured by \citet{1988ApJS...68...91R} and
\citet{1995AJ....110.1115D}. Horizontal error bars for each photometry
point show the width of the used filter. The data have been fitted
(red line) to a model of the emission of the stars (a single burst
with a exponential SFR), the gas (including the most common emission
lines, shown in the plot with a width equal to the resolution of the
stellar model at each wavelength), and the dust (including two dust
components of two different temperatures, and aromatic molecules). Our models of the
dust emission are compared with the ones carried out following the
method in \citet[][green line]{draine06}.}
\end{center}
\end{figure*}

\slugcomment{Please, plot this figure with the width of two columns}
\placefigure{some_fits}

The fitting procedure is based on a Monte-Carlo technique.  With our
assumptions about the components in the observed SED, we searched a
limited set of parameters for dust mass (through $C_{i}$) and
temperature for each component. We assumed temperature ranges of
200--750~K, 35--100~K, and 10--25~K for the aromatics, warm, and cold
silicates, respectively.  At each set of temperature points, a similar
grid of mass scalings ($C_{i}$) was explored for each component.
After an initial, coarse grid search, a finer temperature and mass
scaling grid was defined for each component and the procedure
repeated.  We iterated this procedure (typically only two refinements
of the grid were needed) until a good fit was found based on a
$\chi^2$ goodness of the fit estimator. While the results of this
technique are not unique, it yields representative estimates of both
the dust mass and temperature (and errors obtained through the
analysis of the $\chi^2$ distribution), given the input
assumptions. Some examples of the final fits to the entire SEDs are
shown in Figure~\ref{some_fits}. Note that our simplistic models
(which only assume two dust temperatures) usually produce an
artificial inflexion point at around 50-60~\mic\, (where the cold and
warm dust components join). This artifact is also present in other
models (for example, in the ones plotted in Figure~\ref{some_fits}
using the technique described in
\citealt{draine06}), but it is not seen
in any IR spectrum. However, it does not affect our estimates of the
integrated IR luminosities, given that the energy output is dominated
by emission at longer wavelengths.

The small number of independent observations in the MIR and FIR
justifies the simplistic approach in our models with just three
emitting components that need 5 free parameters to fit. We will use
these models to estimate the integrated IR luminosities of regions in
M81 and to study the characteristics of the dust in those regions. We
tested our results with other more complicated models (that use more
free parameters and/or assumptions) to check that our approach did not
introduce any systematic effects in those areas. We used comparison
models developed by \citet[][see the next Section for more
details]{2002ApJ...576..159D} and \citet{draine06} to characterize all
the SINGS galaxies. The latter models are based on the study of the
Milky Way extinction carried out in
\citet{2001ApJ...548..296W}. Briefly, these authors assume a certain
stellar radiation field (based on the NIR data in our case) that heats
the dust grains. The distribution of sizes of the dust grains is that
observed for the Milky Way. The grain temperatures are calculated
using a complete treatment of temperature fluctuations following
single-photon heating and by studying the thermal equilibrium of the
dust. Figure~\ref{some_fits} shows the comparisons of our fits and the
ones carried out with the models by
\citet{draine06}. The fits are very similar, showing
large differences only for wavelengths longer than 160~\mic, where the
models by \citet{draine06} sometimes overpredict the observed
flux. Despite this, our estimations of the TIR luminosities are very
similar to those obtained from the comparison models (the average
difference is less than 4\%, with a scatter of 6\%). However, the dust
masses we obtain with our models are systematically smaller (on
average, 0.2$\pm$0.2~dex). This difference is linked to the sizes and
composition of the grains assumed by each set of models. While we
assume a fixed dust grain size and two dust temperatures,
\citet{draine06} assume a dust size distribution based on Milky Way
abundances, producing a range of temperatures. This size distribution
includes larger amounts of cold dust than our two component fits. The
difference in dust masses is comparable to the quoted uncertainties
(see Figure~\ref{mass_vs_radial}) and does not affect our results.

\subsection{Degeneracies of the models}

Our stellar population synthesis models are subject to the typical
degeneracies in this kind of study, for example age-metallicity
(\citealt{1994ApJS...95..107W}) or extinction-age (see, e.g.,
\citealt{2002AJ....123.1864G},
\citealt{2003MNRAS.338..508P}, and \citealt{2003MNRAS.341...33K}). 

The well-known age-metallicity degeneracy was artificially removed in
our models by fixing the metallicity to the solar value based on
estimates by several authors
\citep{1994ApJ...420...87Z,1984AJ.....89.1702S,2000AJ....119.2745K}.


The main source of degeneracy in our stellar emission models then
comes from the lack of knowledge about the star formation history of
the galaxy. Indeed, similar SEDs can be obtained with a young
instantaneous ($\tau\sim0$) burst or with older and less intense
continuous star formation ($\tau\gg1$). Therefore, our simplistic
assumption of an exponentially decaying burst of star formation can be
very far from the reality, where several small bursts may have
occurred at different epochs. The final photometry would be, in
general, blind to that formation history. However, given that M81 is
an early-type spiral, it would be expected to have an old and massive
burst forming the bulge, where most of the stars of the galaxy
lie. This old stellar population would dominate the emission in the
NIR. More recent star formation (switching on and off during the life
of the galaxy) would be concentrated in the disk, with the most recent
events contributing significantly to (or even dominating) the UV
emission. Extinction can make the two scenarios previously described
nearly indistinguishable.

However, the extinction and the lack of knowledge about the star
formation history have only a minor effect on our results. In this
paper, the models have been used to estimate the contribution of
stellar light to the fluxes in the IRAC and MIPS bands. At these
wavelengths, the extinction is negligible. Variations in the age or
the star formation history do not strongly affect the shape of the SED
in the NIR, particularly when the emission at these wavelengths is
dominated by old stars (at least, for an early type spiral such as
M81).

In the case of the dust emission models, there are to main
degeneracies. The first comes from the lack of knowledge about the
aromatic spectrum in M81, and about the possibility of changes of that
spectrum from region to region. Given this degeneracy, we will avoid
any discussion about aromatic molecules in this paper. The second
degeneracy is linked to the limited data in the FIR, which is unable
to locate the emission peak accurately. Indeed, the SEDs seem to
continue rising beyond our reddest point at 160~\mic\, (or the maximum
is reached at approximately that wavelength). Based on our models and
the ones developed by \citet{draine06}, we estimate that the lack of
data at $\lambda>160$~\mic\, leads to an uncertainty of less than a
factor of 2 in the derived integrated IR luminosities and dust masses
(which would be underestimated by our models if a significant amount
of cold dust -with $T\lesssim15$~K- is also present in M81). This
uncertainty is smaller than the quoted errors. The absence of any data
in the MIR between 24 and 70~\mic\, should not affect the results
significantly, since the energy in this wavelength regime is a small
fraction of the total in the IR.

\section{PROPERTIES OF THE DUST IN M81}
\label{results_1}

In this Section, we will describe our main results on the properties
of the dust in M81. All the results in this Section were obtained from
the 160RES dataset because the dust emission is better characterized
in this case (FIR data is available), and they span a larger range of
radial distances, mapping most of the galaxy. We analyze the stellar
and dust emission in the MIR from a statistical point of view, and
estimate the integrated IR luminosity, which will be used in the
comparison of SFR estimators in Section~\ref{results_2}. We will
specifically discuss the results obtained with the LOCALBKG
photometry, where the calculated integrated luminosities refer to the
dust heated only by the young stars, since the more diffuse emission
(of more uncertain origin) is removed. We also include a discussion
about the GLOBALBKG photometry, which might be useful for the study of
the global stellar content in sub-galactic regions. In the last part
of this Section, we will concentrate on the radial dependence of the
dust properties in M81. Given that we are interested in the global
characteristics of the dust, the results in the last subsection refer
to the photometry measured with the regions selected at 160RES where
the background was estimated globally (to include all the emission).

\subsection{Stellar emission in the MIR}
\label{ir_prop}

Our stellar population models show that the emission of the regions
selected at 160RES is completely dominated by stars at 3.6~\mic\, and
4.5~\mic, with very little contamination from the dust: the median and
standard deviation of the fraction of the total emission coming from
the dust for all the 160RES regions is $0\pm15$\% at 3.6~\mic\, and
$0\pm10$\% at 4.5~\mic. Stellar emission is also dominant at
5.8~\mic\, ($40\pm20$\% of the flux comes from the dust). At redder
wavelengths, most of the emission comes from the dust: $80\pm20$\% of
the emission at 8~\mic, $93\pm7$\% at 24~\mic, and $100\pm1$\% at 70
and 160~\mic. These fractions are not affected by the intrinsic
degeneracies of the modeling procedure, given that the models in this
part of the spectrum are not very sensitive to changes in age or
metallicity (the SEDs are dominated by the Rayleigh-Jeans spectrum of
cold old stars), and the NIR and MIR are not strongly affected by
extinction.

The fractions of the dust emission in the IRAC bands were checked
using the recipe found in \citet{2004ApJS..154..235P}. The emission at
3.6 and 4.5~\mic\, is assumed to be completely dominated by stars and
we can estimate the contribution of the stellar emission to the 5.8
and 8.0~\mic\, channels using the typical colors of an M0 III star
[$m(3.6)-m(4.5)=-0.15$~mag, $m(4.5)-m(5.8)=+0.11$~mag,
$m(5.8)-m(8.0)=+0.04$~mag in the Vega system, corresponding to
$F_\nu(3.6)/F_\nu(4.5)=1.78$, $F_\nu(4.5)/F_\nu(5.8)=1.39$, and
$F_\nu(5.8)/F_\nu(8.0)=1.78$], which should dominate the stellar
emission at these wavelengths. Using this recipe, we obtain an average
dust contribution to the 5.8 and 8.0~\mic\, emissions of $45\pm26$\%
and $77\pm22$\%, respectively.

Note that the regions used at 160RES enclose both star-forming
complexes and/or zones of the arms where the emission is most probably
dominated by older stars. The size of the photometry apertures has an
effect on the stellar emission fractions previously quoted: for actual
HII regions (i.e., for smaller apertures such as the 24RES ones) where
the emission from young stars and warm dust is relatively more
important, the fractions change by significant factors. According to
our stellar population models, the 3.6~\mic\, and 4.5~\mic\, emissions
have a non-negligible dust component: $10\pm10$\% and $10\pm10$\% of
the total 3.6 and 4.5~\mic\, flux comes from the dust,
respectively. Note that heavily extincted young stars that would be
invisible in the optical could contribute to the observed flux in the
NIR at 24RES. According to our models, the dust emission dominates the
total flux at redder wavelengths: $70\pm20$\% at 5.8~\mic, $90\pm10$\%
at 8.0~\mic, and $100\pm10$\% of the total emission at 24~\mic.

\subsection{Integrated IR emission}
\label{ir_calc}

In this Section, we use the regions selected in the 160RES image set
to investigate possible relationships between monochromatic fluxes and
colors in the MIR and FIR and the total emission of the dust
integrated from 8 to 1000~\mic\, [TIR luminosity, $L(8-1000)$
hereafter]. The following discussion will be mainly based on the
photometry obtained for a GLOBALBKG. We will analyze the effect on
these results of the diffuse IR emission by also using the photometry
obtained with a LOCALBKG. This should give relationships which are
better suited for SFR studies. The contribution of the stellar light
has been subtracted from the MIR-FIR photometry prior to determining
$L(8-1000)$.

\subsubsection{The estimation of $L(8-1000)$}
\label{estimltir}

TIR luminosities were calculated for the 160RES regions by integrating
the fitted dust models. We also calculated the integrated luminosity
from 3 to 1100~\mic\, [$L(3-1100)$] to compare our data with the
semi-empirical models developed by \citet{2002ApJ...576..159D}. These
authors obtained an equation (Equation~4 in that paper) to estimate
the $L(3-1100)$ from the three MIPS fluxes. This relationship is based
on models of the global emission of normal star-forming galaxies with
$L(3-1100)\gtrsim10^{8}\,L_\odot$. Thus, the application to faint
individual HII regions within a galaxy may not be justified. In fact,
the SEDs for sub-galactic regions in M81 are colder than the coldest
model in the \citet{2002ApJ...576..159D} template set, as also noted
by \citet{2005ApJ...633..857D} for sub-galactic regions in M51 and
NGC7331, and by \citet{2004ApJS..154..253H} for NGC300.

For the GLOBALBKG photometry, our own estimations of $L(3-1100)$ are
systematically fainter (7\% on average, with a scatter of 6\%) than
the ones obtained using Equation~4 in
\citet{2002ApJ...576..159D}. The scatter is mainly due to the two
nuclear regions and the coldest regions in the inter-arm zone; without
these regions, the scatter is 2\%. For the entire galaxy, our models
give a value of $L(3-1100)$ that is 1\% lower than the one obtained
with the equation in \citet{2002ApJ...576..159D}. These small
systematic differences (comparable to the uncertainties in the
measured MIPS fluxes) are consistent with the fact that the
sub-galactic regions in M81 are colder than the coldest model in the
\citet{2002ApJ...576..159D} template set.

If we use the photometry calculated with the LOCALBKG, the values of
$L(3-1100)$ estimated from our models are 5-15\% lower than the ones
for the GLOBALBKG. This is mainly caused by a decrease in the
160~\mic\, flux due to the substantial diffuse emission at that
wavelength. The diffuse component accounts on average for 36\% of the
total 8~\mic\, emission of the 160RES regions, 23\% at 24~\mic, 26\%
at 70~\mic, and 34\% at 160~\mic. The average difference between our
luminosities (for the LOCALBKG, i.e., after subtracting the diffuse
emission) and those estimated with the Equation in
\citet{2002ApJ...576..159D} is $10\pm3$\%. In this case, there is a
clear linear correlation where the difference is higher for fainter
regions: up to 15\% for the faintest source at
$L(3-1100)\sim10^{6.8}\,L_\sun$ and 5\% for the brightest at
$L(3-1100)\sim10^{8}\,L_\sun$.

These comparisons suggest that the equations derived for entire
galaxies with high IR luminosities should be corrected by small
amounts when applying them for sub-galactic regions with moderate SFRs
(and for entire galaxies with low IR luminosities). The reason is that
the emission of the cold dust is relatively brighter for pieces of
galaxies or galaxies with moderate SFRs and dust contents, and the
diffuse emission plays an important role. This diffuse emission is
extremely relevant for the wavelengths around 160~\mic, which
contribute the most to the TIR luminosity, as we show in the following
discussion.

\subsubsection{Correlating the three MIPS fluxes to $L(8-1000)$}

Relying on our models of dust emission for the different 160RES
regions in M81, we obtained simple relationships (using a singular
value decomposition method) to obtain $L(8-1000)$ and $L(3-1100)$ from
the fluxes at the three MIPS wavelengths (after subtracting the
stellar component) for sub-galactic regions in the range $10^7\lesssim
L(8-1000)\lesssim 10^{8}$~L$_\sun$. In all the following Equations,
$L(\mathrm{filter})$ refers to $\nu_\mathrm{eff}~L_\nu$ in units of
erg~s$^{-1}$ or solar units, where $\nu_\mathrm{eff}$ is the effective
frequency of the filter. The relations for the GLOBALBKG case, in
solar units, are\footnote{All the correlations in this Section have
been obtained by fitting the data for sub-galactic regions (i.e., not
using the data point for the entire galaxy) and accounting for the
uncertainties.}:



\begin{displaymath}
L(8-1000)=(+1.439\pm0.415)\times L(24)+\\
\end{displaymath}
\begin{equation}
\label{3mips_tir}
+(+0.814\pm0.099)\times L(70)+(+1.1229\pm0.039)\times L(160)
\end{equation}

\begin{displaymath}
L(3-1100)=(+1.832\pm0.420)\times L(24)+\\
\end{displaymath}
\begin{equation}
+(+0.779\pm0.094)\times L(70)+(+1.176\pm0.031)\times L(160)
\end{equation}

For both fits, the data show a scatter of 1\% around the given
relationships.

\begin{figure*}
\begin{center}
\includegraphics[angle=-90,width=7.5cm]{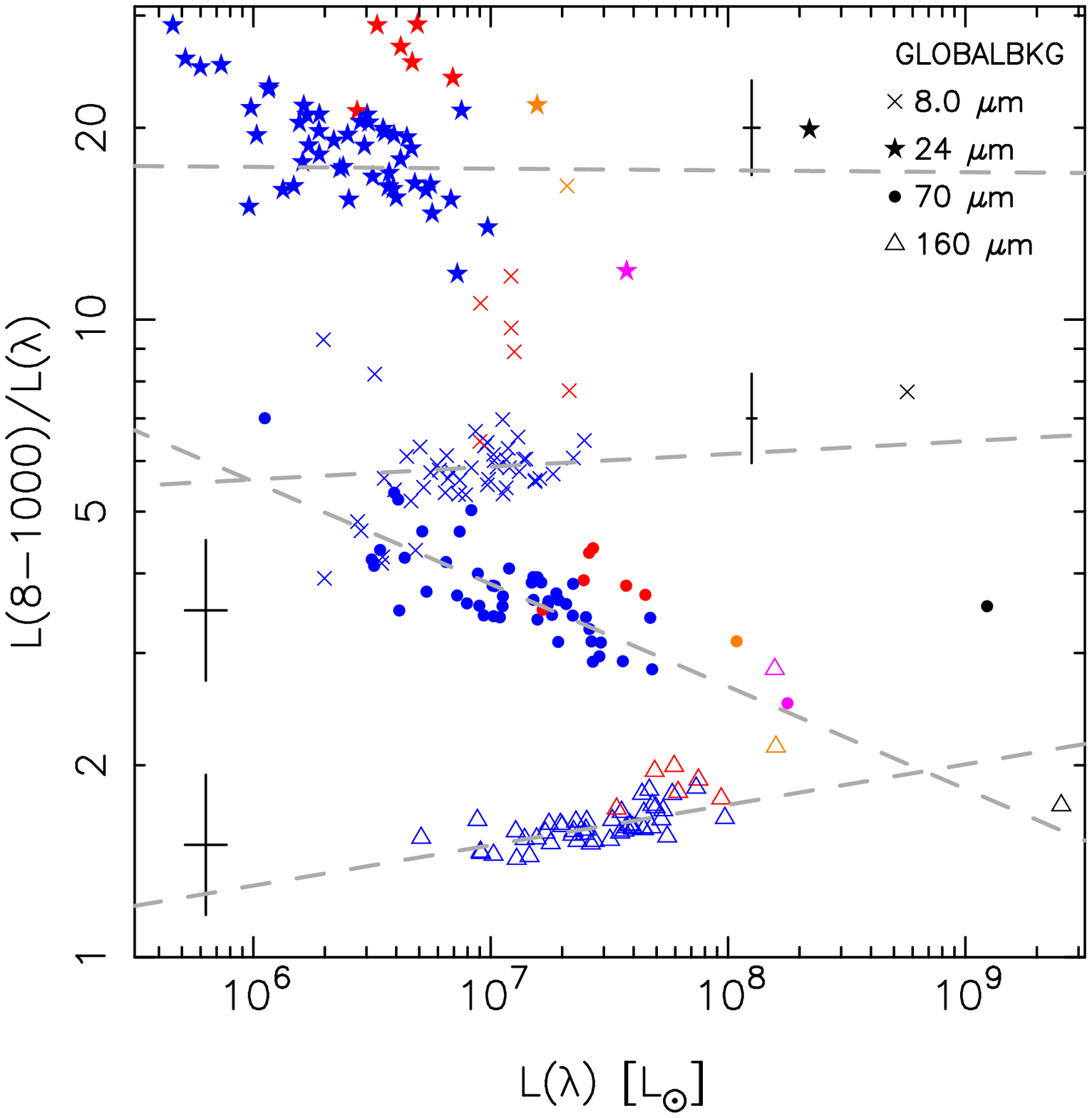}
\hspace{0.5cm}
\includegraphics[angle=-90,width=7.5cm]{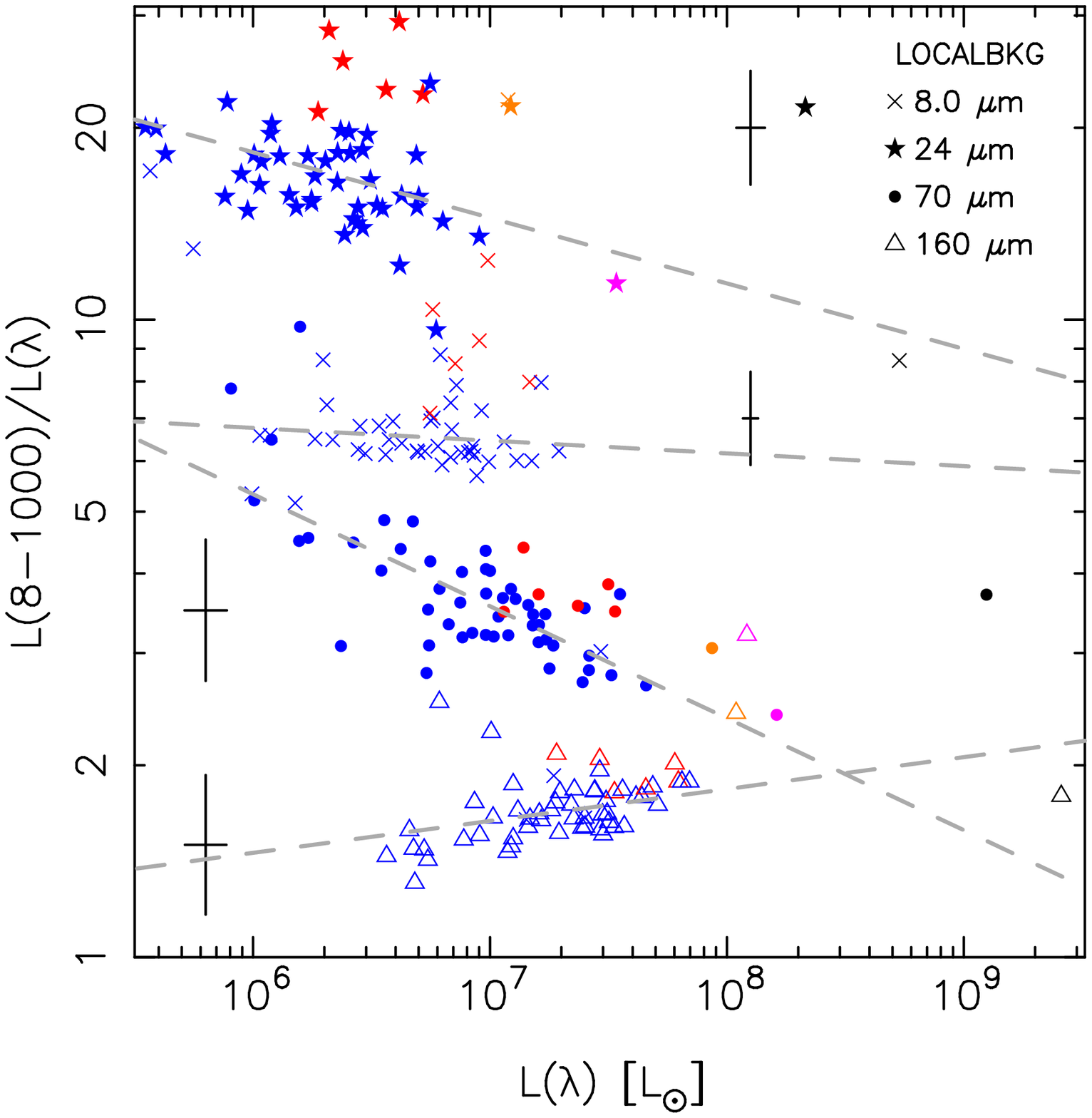}
\figcaption{\label{mono_to_tir} Relationship between the MIR and FIR
monochromatic emissions and the 8-1000~\mic\, luminosities for
star-forming complexes in M81 (selected in 160RES resolution
images). On the left panel, the data refer to the photometry
calculated with a global background (GLOBALBKG case in the text). On
the right, a local background (LOCALBKG case) was used to remove the
diffuse component of the emission in the IR. For both panels, the
integrated luminosities have been calculated by fitting models of dust
emission to the SEDs for wavelengths redder than 5~\mic. The best
linear fits to the data at each wavelength are also shown. The
different colors refer to the distinct types of regions defined in the
left panel of Figure~\ref{reg_160}. In both panels, average errors for
the different luminosities (horizontal bars) and luminosity ratios
(vertical bars) are plotted to one side of the corresponding data
points.}
\end{center}
\end{figure*}

\slugcomment{Please, plot this figure with the width of one column}
\placefigure{mono_to_tir}

The same equations for luminosities estimated after subtracting the
diffuse emission (LOCALBKG case) are:



\begin{displaymath}
L(8-1000)=(+0.202\pm0.510)\times L(24)+\\
\end{displaymath}
\begin{equation}
\label{3mips_tir_nodiffuse}
+(+0.802\pm0.160)\times L(70)+(+1.303\pm0.061)\times L(160)
\end{equation}

\begin{displaymath}
L(3-1100)=(+0.294\pm0.551)\times L(24)+\\
\end{displaymath}
\begin{equation}
+(+0.766\pm0.190)\times L(70)+(+1.392\pm0.063)\times L(160)
\end{equation}

The scatter of the data points around these relations is approximately
4\% in both cases.

\subsubsection{Correlating monochromatic MIR fluxes to $L(8-1000)$}

Given the poorer resolution of the MIPS 70 and 160~\mic\, channels
relative to the 24~\mic\, channel, it would be very convenient to use
the 24~\mic\, emission alone to obtain reliable IR luminosities. This
would allow us to study the star-forming complexes in M81 (or any
other resolved galaxy) at high enough angular resolution to resolve
individual HII regions and compare the IR results with other SFR
estimators, such as H$\alpha$ or the UV.

The degree of reliability of using only one monochromatic flux to
estimate the 8-1000~\mic\, luminosity is shown in
Figure~\ref{mono_to_tir}, including the diffuse emission at all
wavelengths (left panel) and removing it (right panel). Here we plot
the TIR luminosity of each individual region estimated with the three
MIPS fluxes (after removing the stellar component) as a function of
the monochromatic luminosity in those MIPS bands. The best fits to the
GLOBALBKG data are the following (with all luminosities in solar
units):

\begin{displaymath}
\log{[L(8-1000)]}=(+0.63\pm0.11)+
\end{displaymath}
\begin{equation}
+(+1.020\pm0.015)\times\log{[L(8.0)]}
\end{equation}

\begin{displaymath}
\log{[L(8-1000)]}=(+1.25\pm0.06)+
\end{displaymath}
\begin{equation}
\label{24_tir}
+(+0.997\pm0.010)\times\log{[L(24)]}
\end{equation}

\begin{displaymath}
\log{[L(8-1000)]}=(+1.71\pm0.46)+
\end{displaymath}
\begin{equation}
+(+0.839\pm0.065)\times\log{[L(70)]}
\end{equation}

\begin{displaymath}
\log{[L(8-1000)]}=(-0.27\pm0.36)+
\end{displaymath}
\begin{equation}
+(+1.064\pm0.048)\times\log{[L(160)]}
\end{equation}

The diffuse emission has a very significant effect on the previous
correlations, especially for the 8 and 160~\mic\, data. The
relationships between the monochromatic fluxes and the TIR luminosity
for the photometry where the diffuse emission has been removed
(LOCALBKG case) are:

\begin{displaymath}
\log{[L(8-1000)]}=(+0.95\pm0.13)+
\end{displaymath}
\begin{equation}
+(+0.980\pm0.019)\times\log{[L(8.0)]}
\end{equation}

\begin{displaymath}
\log{[L(8-1000)]}=(+1.88\pm0.27)+
\end{displaymath}
\begin{equation}
\label{24_tir_nodiffuse}
+(+0.897\pm0.041)\times\log{[L(24)]}
\end{equation}

\begin{displaymath}
\log{[L(8-1000)]}=(+1.78\pm0.27)+
\end{displaymath}
\begin{equation}
+(+0.824\pm0.038)\times\log{[L(70)]}
\end{equation}

\begin{displaymath}
\log{[L(8-1000)]}=(-0.14\pm0.30)+
\end{displaymath}
\begin{equation}
+(+1.050\pm0.041)\times\log{[L(160)]}
\end{equation}

Figure~\ref{mono_to_tir} clearly shows that the best estimator of the
TIR luminosity (the one presenting the lowest scatter) is the
160~\mic\, emission, which accounts for more than 50\% of the TIR
luminosity (see also Equations~\ref{3mips_tir} and
\ref{3mips_tir_nodiffuse}). The scatter around a linear relationship
for the monochromatic 160~\mic\, data is 5\% of the TIR luminosity
(6\% if we remove the diffuse emission). The 24~\mic\, data alone is a
much poorer estimator of the TIR luminosity, presenting a scatter
around the given relationship of 12\% (13\%). The goodness of the
70~\mic\, and the 8~\mic\, data as estimators of the TIR luminosity
lies in between: 9\% (11\%) and 8\% (7\%) scatter around the given
relationships, respectively. It is important to notice that, although
8.0~\mic\, seems to be a better TIR luminosity estimator than
24~\mic\, based on the scatters of the fits, the IRAC band is more
affected by the stellar emission contamination (more than a factor of
2 difference). Moreover, the 8.0~\mic\, image shows a brighter diffuse
emission component than the 24~\mic\, image, 36\% of the total
emission for the former and 23\% for the latter
(cf. Section~\ref{estimltir}). According to
\citet{2004ApJS..154...10F}, the scattered light in the 8~\mic\, array
can only account for $\sim$2\% of that fraction. This clearly shows
that the 8~\mic\, emission depends not only on the local star
formation rate but also in some other mechanisms related to the
formation and destruction of aromatic molecules
\citep[see][and references therein]{2005ApJ...633..871C}. Indeed, in
the LOCALBKG case (where the starlight contamination and the diffuse
emission of uncertain origin should be removed), the correlation
between the 8~\mic\, and the TIR luminosities improves.

Figure~\ref{mono_to_tir} also shows one interesting point about the
contribution of the various wavelengths to the integrated
luminosity. As we move to higher TIR luminosities, the contribution of
the 160~\mic\, luminosity to the integral becomes less important (as
the positive slope shows), while the contribution of the hotter dust
emitting at shorter wavelengths rises (as shown by the negative slopes
of the linear fits for 24 and 70~\mic). Figure~\ref{mono_to_tir} shows
that the dust contributing the most to $L(8-1000)$ has a temperature
of around $T\sim19$~K (its emission peaks at $\lambda\sim160$~\mic)
for $L(8-1000)\lesssim10^8\,L_\sun$. The fraction of $L(8-1000)$
emitted by the dust at $T\gtrsim40$~K (with emission peaking at
$\lambda<70$~\mic) increases as $L(8-1000)$ gets brighter (i.e., as
the current SFR increases). The hotter dust contributes as much as the
colder dust component at $L(8-1000)\gtrsim10^{8.5}\,L_\sun$ (i.e., for
$SFR\gtrsim$0.05~$\mathcal{M}_\sun\,\mathrm{yr}^{-1}$). The fraction
of $L(8-1000)$ emitted by dust at $T\gtrsim125$~K (with emission
peaking at 24~\mic) is always lower than 10\% for
$L(8-1000)\lesssim10^{9.0}\,L_\sun$. Another interesting point in
Figure~\ref{mono_to_tir} is that the regions dominated by diffuse
emission (red symbols) seem to deviate significantly from the
star-forming complexes (blue symbols), showing a larger contribution
to $L(8-1000)$ from the colder dust.

Finally, it is also worth noticing that the entire galaxy and the
nuclear and circumnuclear zones deviate significantly from the
relationship found for the star-forming regions in the arms of M81.
These regions present systematically hotter dust temperatures than the
ones obtained for the zones in the arms (see Section~\ref{dust_prop}),
which should be linked to other sources of heating, such as the
central AGN, the bulge old stars, or dynamical friction
\citep{gordon06}.

\begin{figure}
\begin{center}
\includegraphics[angle=-90,width=9cm]{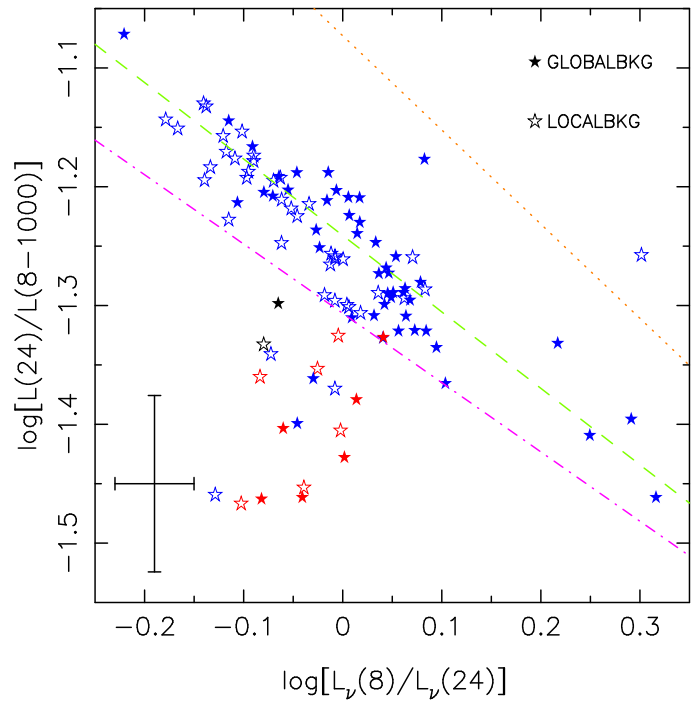}
\figcaption{\label{make_24_better}The 24-to-TIR luminosity ratio as a function 
of the 8.0-to-24~\mic\, color for different regions selected in the
160RES image set. Filled stars show the results for the photometry
calculated with a global background, and open stars for the local
background. From top to bottom, the linear relationships show the
correlations found by \citet{2005ApJ...633..871C} for 21 regions in
M51 (dotted line), our own correlation (dashed line), averaged from
Equations~\ref{24_8_tir_1} and \ref{24_8_tir_2}, and the prediction
from the models of \citet[][dash-dotted line]
{2002ApJ...576..159D}. The different colors refer to the distinct
types of regions defined in the left panel of Figure~\ref{reg_160}.}
\end{center}
\end{figure}

\slugcomment{Please, plot this figure with the width of one column}
\placefigure{make_24_better}

\subsubsection{Correlating MIR colors to $L(8-1000)$}

In Figure~\ref{make_24_better}, we explore the possibility of using a
color to improve the relationship between the 24~\mic\, and the TIR
luminosities. In principle, we could expect a color involving MIR
fluxes to be correlated with the behavior of the SED in the FIR, i.e.,
the amount and emissivity of the cold dust (dominating the FIR
luminosity) might be linked with the properties of the warm dust if
both dust components share the same heating source. If that were the
case, we would be able to reduce the scatter in the estimation of
$L(8-1000)$ from single MIR fluxes. Figure~\ref{make_24_better} shows
that there is a correlation between the 8.0~\mic\,-to-24~\mic\, color
and the TIR luminosity, although the scatter is considerable. The best
fit to the data (GLOBALBKG case) gives:

\begin{displaymath}
\log{[L(8-1000)/L(24)]}=(+0.94\pm0.05)+
\end{displaymath}
\begin{equation}
\label{24_8_tir_1}
+(-0.644\pm0.102)\times\log{[L(24)/L(8.0)]}
\end{equation}

For the data where the diffuse component has been removed (LOCALBKG
case), the correlation is very similar:

\begin{displaymath}
\log{[L(8-1000)/L(24)]}=(+0.93\pm0.06)+
\end{displaymath}
\begin{equation}
\label{24_8_tir_2}
+(-0.695\pm0.155)\times\log{[L(24)/L(8.0)]}
\end{equation}

Using this relationship, we are able to reduce the scatter in the
determination of $L(8-1000)$ to less than 6\% (11\% for the photometry
with no diffuse emission), i.e., the estimates are two times better
than the ones obtained using 24~\mic\, alone. However, the correlation
has a significant number of outliers, most of them corresponding to
regions where the star formation is not very high and the diffuse
emission is very important (inter-arm regions and zones between the
arms and the nucleus). Our correlation lies very near the predictions
from the models of \citet{2002ApJ...576..159D}, and it is
significantly different from the relationship found by
\citet{2005ApJ...633..871C} for 21 regions in M51, although the slope 
is very similar. These differences must arise from the fact that the
8-to-24~\mic\, color depends on the strength of the aromatic emission
(contributing mainly at 8~\mic), which presents significant variations
from galaxy to galaxy depending on parameters such as the abundance of
aromatic molecules, composition (linked to the metallicity, see
\citealt{2005ApJ...628L..29E}), or temperature of the galaxy
\citep[see][]{2005ApJ...633..857D}. Therefore, one should be cautious in 
using Equations~\ref{24_8_tir_1} and \ref{24_8_tir_2} for any galaxy
before it is checked for a larger sample.

All the correlations found in this Section are summarized in
Table~\ref{IRintegr_corrs}.

\subsection{Spatially resolved properties of the dust in M81}

\label{dust_prop}
\begin{figure*}
\begin{center}
\includegraphics[width=7.5cm,angle=-90]{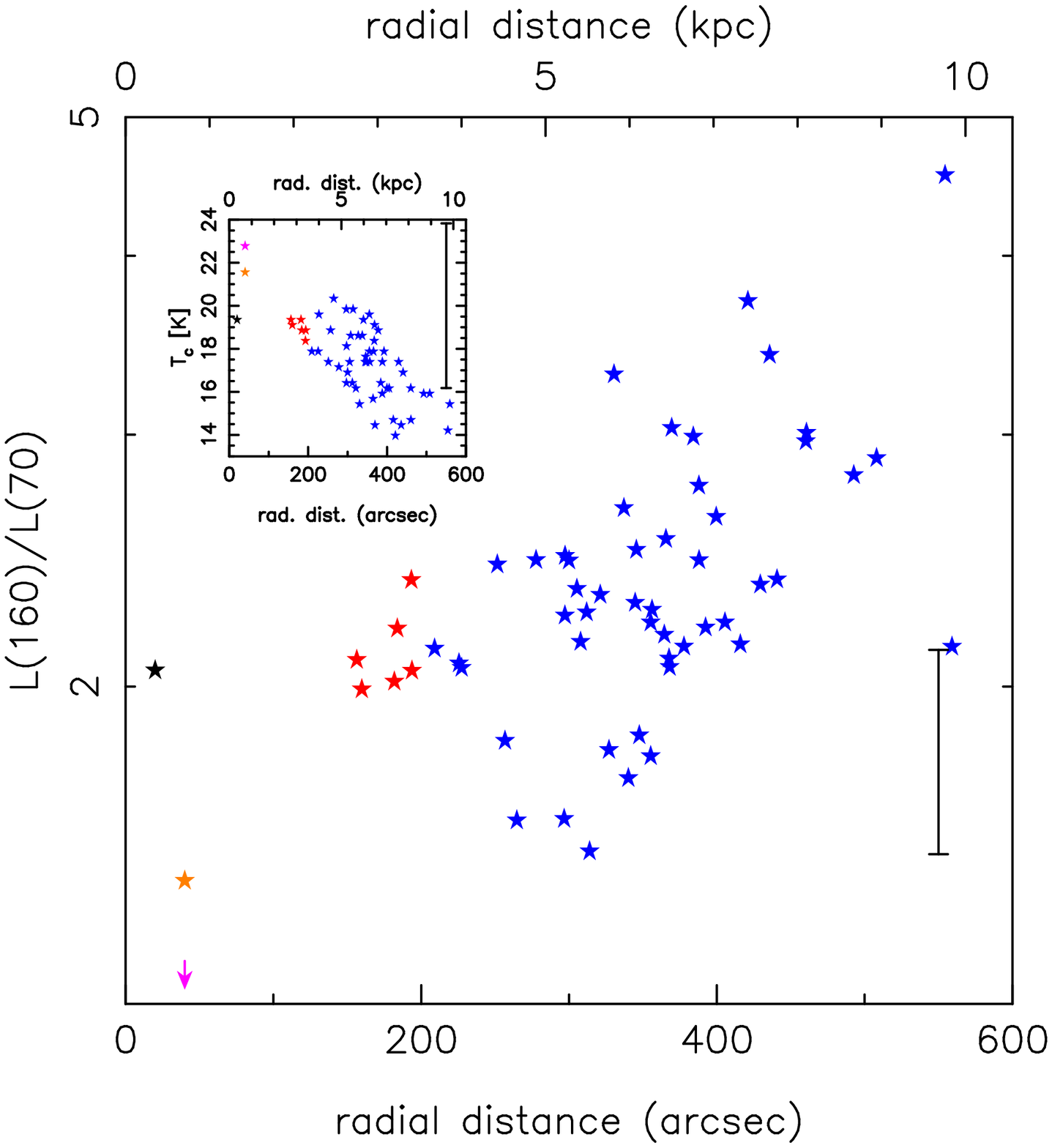}
\hspace{0.2cm}
\includegraphics[width=7.5cm,angle=-90]{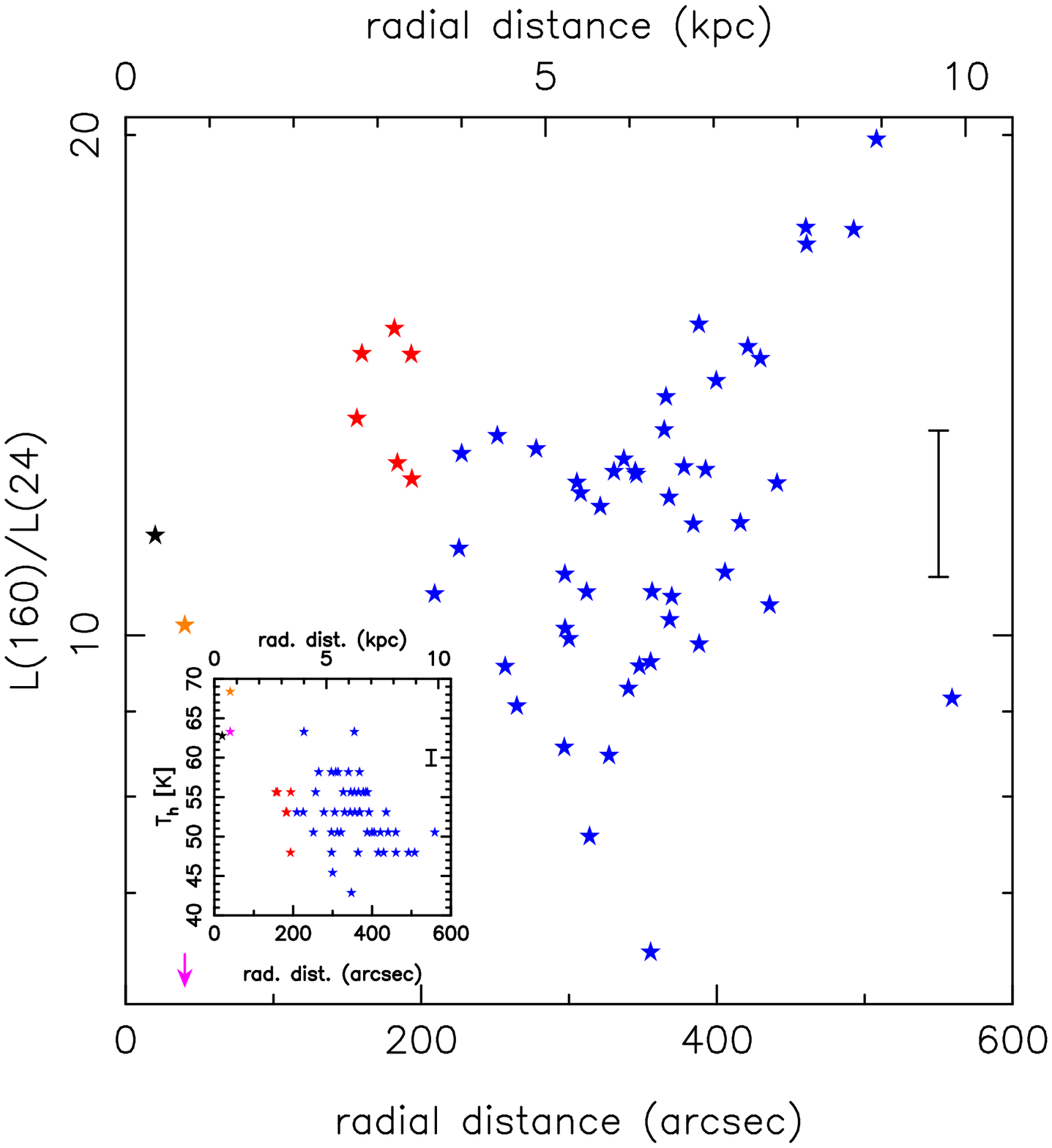}
\figcaption{\label{temp_vs_radial} Radial plots (using de-projected 
radial distances) of the MIPS colors for the regions selected at
160RES. The different colors refer to the distinct types of regions
defined in the left panel of Figure~\ref{reg_160}.  The average
surface density of the entire galaxy has been plotted at
$r=20\arcsec$, and the surface density for the nuclear and the
circumnuclear regions has been plotted at $r=40\arcsec$ (slightly
shifted from $r=0$ for clarity). The insets in each panel show the
dust temperatures for the cold (left panel) and warm (right panel)
components in our models. Average errors for the colors and
temperatures are shown in each panel.}
\end{center}
\end{figure*}

\slugcomment{Please, plot this figure with the width of two column}
\placefigure{temp_vs_radial}

Figure~\ref{temp_vs_radial} shows the MIPS colors as a function of
radius (de-projected radial distances) for the regions selected from
the 160RES dataset\footnote{The results in this Subsection refer to
the GLOBALBKG case, since we are studying the properties of all the
dust emission in M81, including the diffuse component.}. In our
models, the radial dependence of the MIPS colors manifests as a radial
gradient in the dust temperature. The model dust temperatures as a
function of radius are plotted in the insets of
Figure~\ref{temp_vs_radial}. This figure shows that the IR emission in
M81 is dominated by a cold dust component having temperatures around
18~K (the average and standard deviation are $<T_c>=18\pm2$~K). The
warmer dust component has an average temperature of
$<T_w>=53\pm7$~K. In comparison with these temperatures,
\citet{1995AJ....110.1115D} obtain an average dust temperature of
$T\sim30$~K (approximately the average of our two components) for the
entire galaxy and just using the IRAS 60 and 100~\mic\, fluxes and
models with a single dust component. \citet{1995AJ....110.1115D} also
provide expected values for the temperature profiles of the dust. The
temperatures in these profiles range from $T\sim15$~K to $T~\sim50$~K
(depending on the composition of the dust), values that are very
similar to the ones obtained for our fits. Our dust temperatures are
also comparable to those estimated from other work based on FIR and
sub-mm data, which find cold dust temperatures of $15-25$~K and warm
dust temperatures of $40-60$~K
\citep[][]{2000MNRAS.315..115D,2001MNRAS.327..697D,
2002ApJ...567..221P,2003AJ....125.2361B,2004ApJS..154..204R,
2004ApJS..154..248E}

\input{tab3_2c}

The cold dust component shows a clear trend of decreasing temperature
with increasing radius, although the scatter and fitting errors are
large. The temperature profile is consistent with the observed radial
gradient of the $L(160)/L(70)$ color, which increases from
$L(160)/L(70)\sim2$ to $L(160)/L(70)=3-4$ within the disk. For the warm
dust component, the temperature radial gradient is very small,
resulting from a change in the $L(160)/L(24)$ color of a factor of 2
from the center to the outskirts of M81. The central regions of M81
show slightly hotter dust temperatures in the cold component:
$T_c=22.8\pm1.0$~K and $T_c=21.6\pm0.5$ for the nuclear and
circumnuclear regions, respectively. As we mentioned earlier, this
could be due to the presence of another efficient source of dust
heating, apart from the young stars: an AGN for the nucleus and/or
older stars (whose density is higher in the bulge of the galaxy and
may present radial changes in the stellar SEDs) for both zones
\citep[see][and references therein]{2000ARA&A..38..761G}. These 
effects could also be responsible for the radial gradient in the cold
dust temperatures.

\begin{figure}
\begin{center}
\includegraphics[angle=-90,width=9cm]{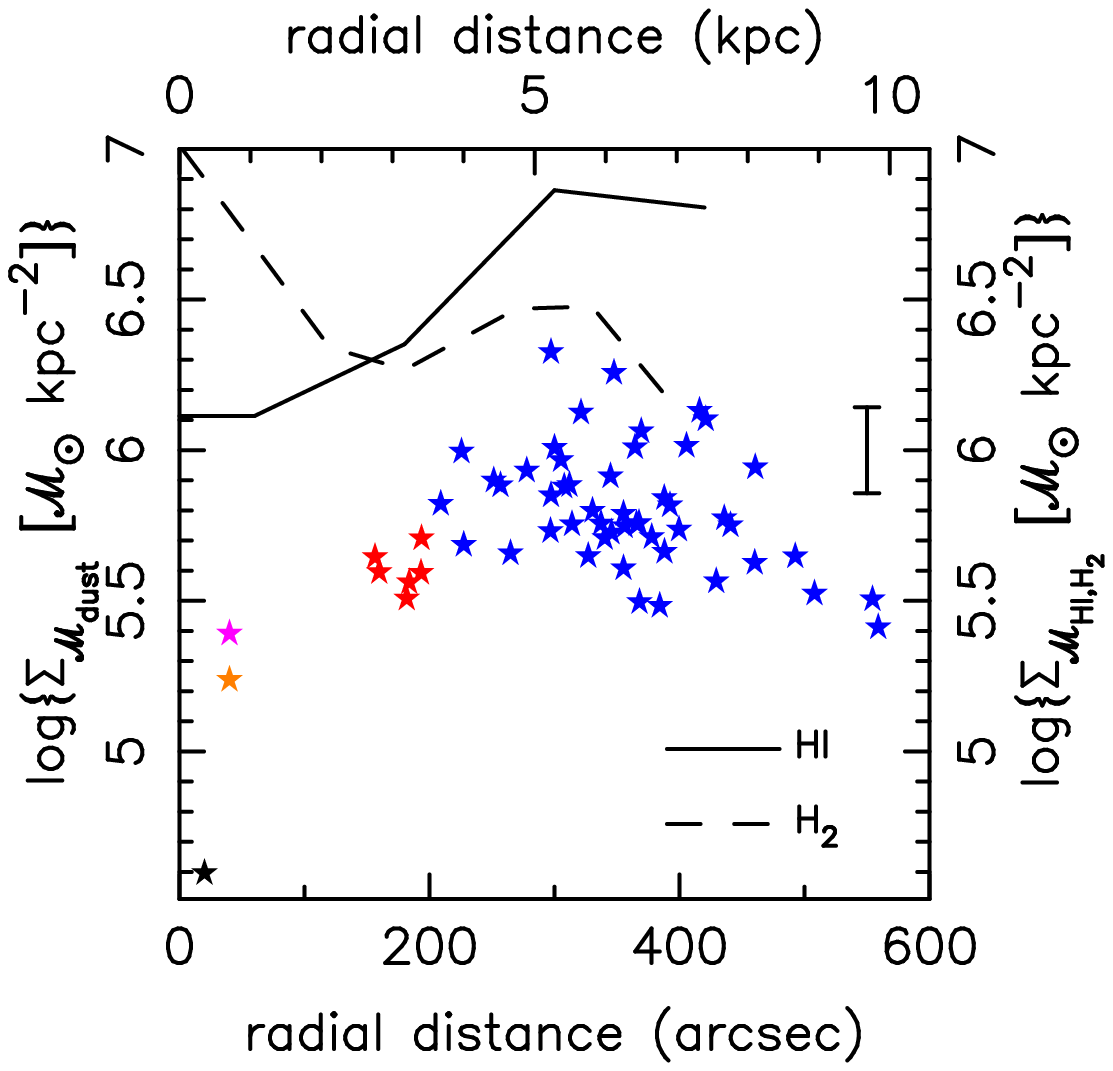}
\figcaption{\label{mass_vs_radial}Radial profile (using
de-projected radial distances) of the dust mass surface density (also
corrected for inclination). The different colors refer to the distinct
types of regions defined in the left panel of
Figure~\ref{reg_160}. The average surface density of the entire galaxy
has been plotted at $r=20\arcsec$, and the surface density for the
nuclear and the circumnuclear regions has been plotted at
$r=40\arcsec$ (slightly shifted from $r=0$ for clarity). The profiles
through the major axis for the mass surface density of the atomic gas
(continuous line) and molecular gas
\citep[dashed line, ][]{1995AJ....110.1115D} are also plotted, with the 
scale shown on the right vertical axis.}
\end{center}
\end{figure}

\slugcomment{Please, plot this figure with the width of one column}
\placefigure{mass_vs_radial}

In combination with the radial gradient of the dust temperature, our
models also reveal a decrease in the dust mass surface density as we
move to higher radial distances (beyond $\sim$300$\arcsec$). This is
shown in Figure~\ref{mass_vs_radial}, where we use de-projected radial
distances and surface densities. This figure also shows the profile
through the major axis of M81 for the mass surface density of the
atomic and molecular gas (estimated by
\citealt{1995AJ....110.1115D} using the data in \citealt{1975ApJ...195...23G} 
and \citealt{1975A&A....45...25R} for the atomic gas, and using the
$^{12}$CO intensities measured by \citealt{1991A&A...242...35B} and
\citealt{1991A&A...242..371S}). For the entire galaxy, the dust-to-gas 
ratio is 0.005 (accounting for both atomic and molecular gas). We
detect a factor of 10 increase in the dust mass surface density from
the center of the galaxy to the inner part of the arms (at a radial
distance of approximately 200$\arcsec$). The atomic gas surface
density also increases by a similar amount in the same region. Within
the arms (from $\sim$200$\arcsec$ to $\sim$400$\arcsec$), the dust
mass density is about 10$^{5.9}$~$\mathcal{M}_\sun\,\mathrm{kpc}^{-2}$
(with a scatter of about 0.2dex). The regions with the highest dust
mass surface densities are found at radial distances of
$300-350\arcsec$, just where the profiles for the molecular and atomic
gas surface densities peak. The dust mass surface density decreases
(by roughly a factor of 10) from $\sim$300$\arcsec$ to the outer
regions. This gradient in the dust density should produce a radial
decrease in the extinction of the stellar light, which is confirmed in
Section~\ref{results_2} (see Figure~\ref{radial_grad}).

One caveat to our results on the radial gradient of the dust mass
density arises from the lack of data at wavelengths longer than
160~\mic\footnote{The only data available for M81 in the sub-mm is a
JCMT/SCUBA observation of the central 2$\arcmin$ of the galaxy which do not
cover any of the star-forming complexes in the arms.}, where the peak
of the SED must be located. Our data are not sufficient to accurately
delimit the position of that maximum, which results in relatively
large uncertainties on the amount of dust and emission arising from
dust colder than approximately 20~K. This means that our estimations
of the dust mass should be considered as lower limits. More data in
the sub-millimeter range (at least, at wavelengths redder than
$\sim$200~\mic, the range that will be covered by Herschel) are
necessary to account for the coldest dust component and robustly
estimate the total amount of dust.


\section{COMPARISON OF STAR FORMATION RATE ESTIMATORS }
\label{results_2}

\begin{figure*}
\begin{center}
\includegraphics[angle=-90,width=7.5cm]{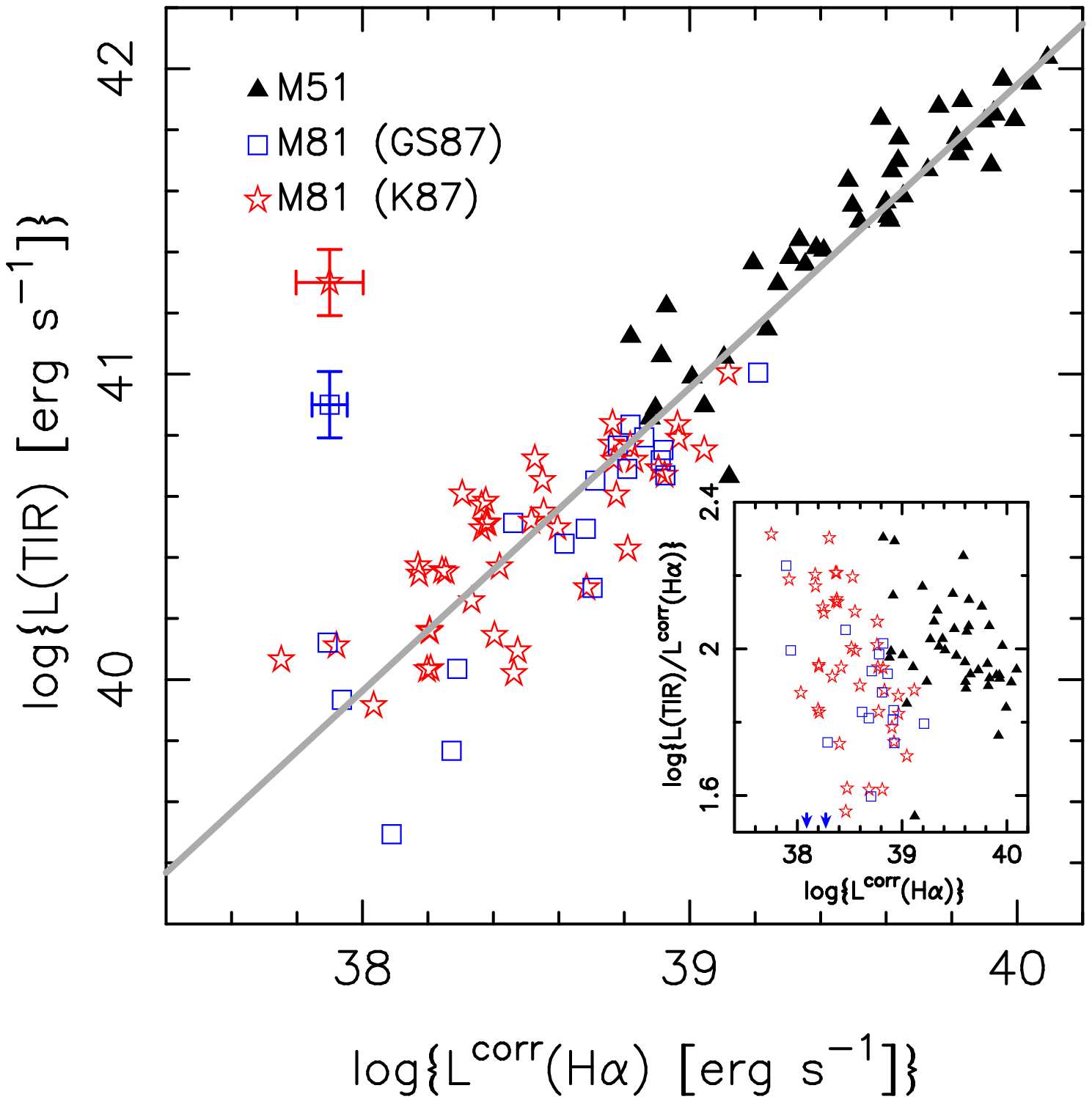}
\hspace{0.2cm}
\includegraphics[angle=-90,width=7.5cm]{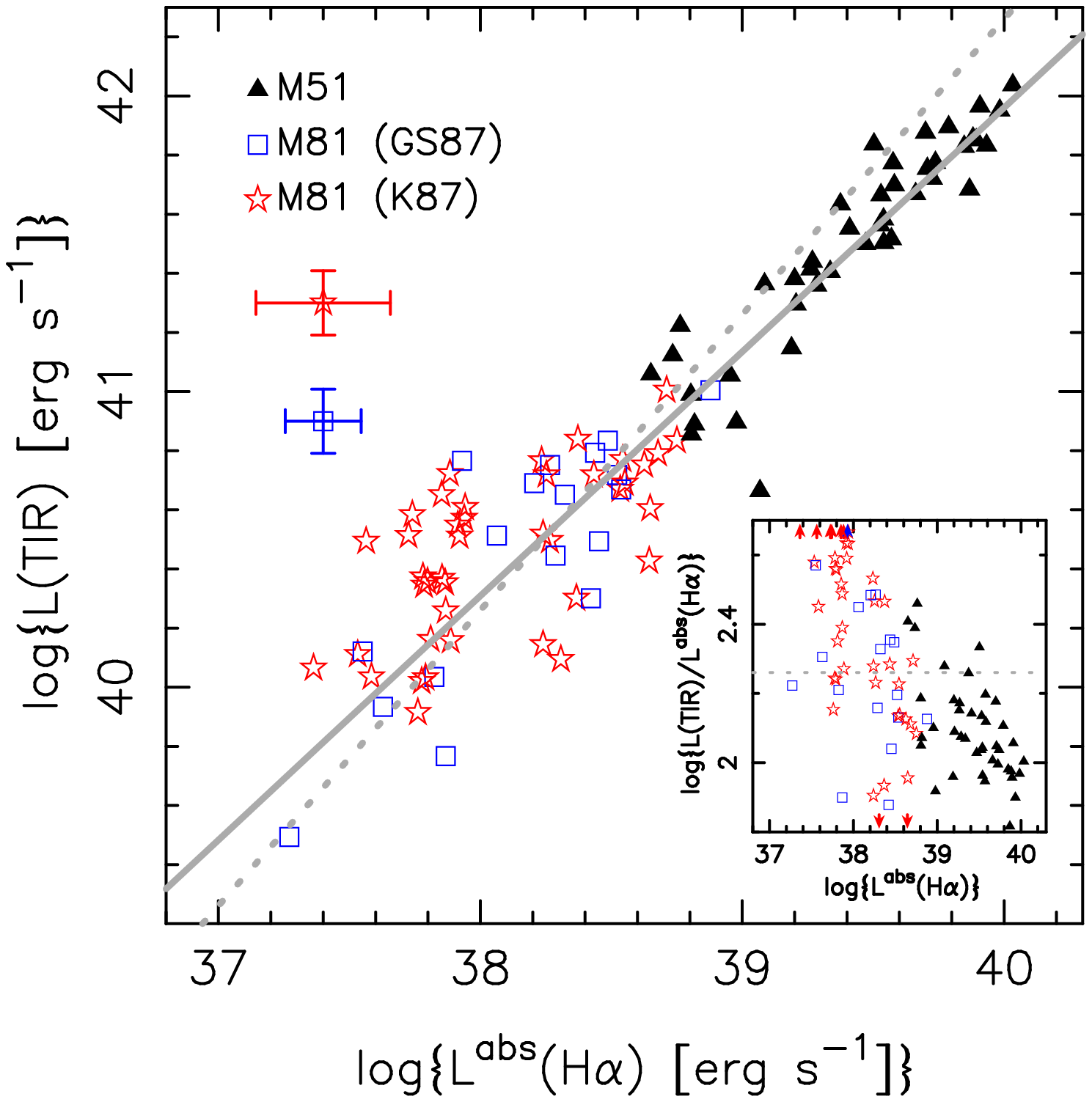}
\figcaption{\label{tir_halpha}On the left, comparison between the 
extinction corrected H$\alpha$ and the TIR luminosities for selected
HII regions in M81 (with different colors and symbols referring to the
distinct types of regions defined in the right panel of
Figure~\ref{reg_160}, extracted from \citealt{1987ApJ...317...82G}
-GS87- and \citealt{1987ApJ...319...61K} -K87-) and in M51 (black
triangles; data extracted from \citealt{2005ApJ...633..871C}). On the
right, the TIR luminosity of the same sample of HII regions is
compared with the H$\alpha$ luminosity absorbed by the dust. For both
panels, we show the typical uncertainties linked to each catalog of
HII regions in M81. The best linear fit is plotted with a continuous
gray line. The insets show the same plots but using a luminosity ratio
in the vertical axis to enhance the differences in the trends followed
by the datasets for each galaxy. The dotted line in the right panel
(and the inset) shows the luminosity relation for slope unity and a
ratio between the H$\alpha$ and TIR luminosities given by the SFR
calibrations found in
\citet{1998ARA&A..36..189K}.}
\end{center}
\end{figure*}

\slugcomment{Please, plot this figure with the width of one column}
\placefigure{tir_halpha}

In this Section, we compare different SFR tracers for 59 individual
HII regions in M81 with available extinction measurements. All of
these HII regions are resolved in our 24RES image set (see the right
panel of Figure~\ref{reg_160}). The sample of 59 sources includes the
18 HII regions in the disk and outskirts of M81 (at radial distances
between 3 and 15~kpc) for which \citet{1987ApJ...317...82G} obtained
optical spectra. These authors measured Balmer decrements that,
appropriately corrected for stellar absorption and temperature
effects, provide a good estimate of the extinction of the gas
emission. The remaining 42 sources in our sample are giant HII regions
(at radial distances between 3 and 10~kpc) for which
\citet{1987ApJ...319...61K} obtained H$\alpha$ and radio (at 6 and
20~cm) observations, estimating optical extinctions for most of
them. These extinction estimations may be affected by radio
synchrotron emission not directly linked to the recent star formation
\citep{1992ARA&A..30..575C}.


In the following, we compare the luminosities of the HII regions in
M81 for 5 SFR tracers: the UV emission, the H$\alpha$ nebular
emission, and 3 luminosities in the IR wavelength range [$L(8-1000)$,
$L(24)$, and $L(8)$, all referring to the photometry measured with a
LOCALBKG]. In this comparison of SFR estimators, we also include the
results obtained by
\citet{2005ApJ...633..871C} for the central region of M51. In that
work, H$\alpha$ and Pa$\alpha$ narrow-band fluxes, jointly with IRAC
and MIPS emissions, are measured for HII regions with high extinctions
($1<A(V)<4$). The combination of the data for M81 and M51 allows us to
probe a wide range of attenuations (from $A(V)\sim0$ to $A(V)\sim4$)
and SFRs (from $3\times10^{-4}$ to
0.1~$\mathcal{M}_\sun\,\mathrm{yr}^{-1}$, or $5\times10^{-3}$ to
0.3~$\mathcal{M}_\sun\,\mathrm{yr}^{-1}\mathrm{kpc}^{-2}$).

The extinction corrected H$\alpha$ luminosities used in this Section
have been estimated by adopting the following differential values
between the H$\alpha$, H$\beta$, and Pa$\alpha$ attenuation factors,
$k(\lambda)$\footnote{The attenuation at a given wavelength,
$A(\lambda)$, is given by $A(\lambda)=k(\lambda)\times E(B-V)$}:
$k(\mathrm{H}\beta)-k(\mathrm{H}\alpha)=1.23$ and
$k(\mathrm{H}\alpha)-k(\mathrm{Pa}\alpha)=2.02$, with
$k(\mathrm{H}\alpha)=2.41$. These factors correspond to a screen
attenuation law with $A(V)\sim0.5$ for M81 and $A(V)\sim3.0$ for M51,
homogeneous distribution of the dust, and dust properties typical of
the Small Magellanic Cloud
\citep{2000ApJ...528..799W}. These attenuation laws seem to best
reproduce the observations for M81 (see, e.g.,
\citealt{1995ApJ...438..181H}) and M51 (see \citealt{2005ApJ...633..871C}). 
Other attenuation laws in \citet{2000ApJ...528..799W} with different
star-dust geometries and distributions, or extinction laws such as the
ones in \citet{1989ApJ...345..245C}, provide similar factors
(different by less than 1\%).  We have assumed the theoretical line
ratios H$\alpha$/H$\beta=$2.86 and H$\alpha$/Pa$\alpha$=8.46, values
expected for a low density gas ($n_{\mathrm{e}}=10^{2}\,\mathrm
cm^{-3}$) with $T_{\mathrm{e}}=10^{4}\,\mathrm K$ in the recombination
Case B including collisional effects \citep{1989agna.book.....O}.

\subsection{The TIR luminosity as a SFR tracer}

The left panel of Figure~\ref{tir_halpha} shows the comparison between
the TIR luminosity (calculated with Equation~\ref{24_8_tir_2}) and the
extinction corrected H$\alpha$ emission for the HII regions in M81 and
M51. The best linear fit to the data for both galaxies is\footnote{All
the linear correlations in this Section will be given for luminosities
in units of erg\,s$^{-1}$.}:

\begin{displaymath}
\log\{L(8-1000)\}=\\
\end{displaymath}
\begin{equation}
\label{tir_hac}
(+2.3\pm1.2)+(+0.992\pm0.031)\times \log\{L^\mathrm{corr}(H\alpha)\}
\end{equation}

The scatter of the data around this relationship is 30\%. This
equation is very different from the one obtained by
\citet{2005ApJ...633..871C} for the M51 regions alone. In fact,
Figure~\ref{tir_halpha} (see the inset in the left panel) shows that
the M81 and M51 data points seem to follow different trends. The
slopes obtained for the M81 or the M51 data alone are very similar
($\sim0.9$), but there is an offset in the zero point (shown in the
inset by the offset in the cloud of points for both galaxies). For the
same value of $L^\mathrm{corr}(H\alpha)$, the M51 regions are brighter
than the M81 ones in the IR. This offset is consistent with the larger
$A(V)$ values found for the M51 central region in comparison with the
extinctions found for the HII regions in the arms of M81. For the
regions with larger extinctions (the ones in M51), more energy is
absorbed by the dust and reemitted in the IR.

Classically, total SFRs for galaxies (or pieces of galaxies) have been
obtained in two different ways. One option is using an estimator
affected by dust attenuation, such as an emission-line or the UV
emission, and correcting the observed flux for the effects of
extinction. Another possibility is directly using a SFR tracer that is
not affected by dust attenuation, such as the IR or radio
emission. Both ways of estimating total SFRs were compared in the left
panel of Figure~\ref{tir_halpha} (where we compare
extinction-corrected H$\alpha$ luminosities with TIR luminosities). In
principle, the IR emission is linked to the dust heated by the light
coming from the newly-formed stars in the HII regions. The UV and
H$\alpha$ emissions (without any extinction correction) are linked to
the photons arising from those young stars that do not interact with
the dust at all. This means that we can obtain an estimate of the SFR
from the observed H$\alpha$ or the UV emission that cannot be
detected by the IR because these photons do not interact with the
dust, and on the other hand, we can account for the emission-line or
UV photons which are extinguished by dust by using IR
observations. Consequently, an alternative way to obtain a total SFR
would be to add the 'unobscured SFR' derived in the UV/optical
(without any extinction correction) and the 'obscured SFR' obtained
from the IR luminosity \citep[see][]{kennicutt06}. This total SFR
should be the same as the SFR derived with the UV/optical estimators
after correcting them for the effects of extinction and taking into
account the spatial variations of the dust content in star-forming
regions, which might result in the UV/optical and IR emissions arising
from different regions (with different attenuations). It should also
coincide with the SFR derived with the IR emission alone, if we can
obtain a calibration for it that takes into account the amount of
star formation not obscured by dust (or if this star formation is
negligible in comparison with the star formation hidden by dust).

The alternative method of obtaining the total SFR adding the
estimations of the unobscured SFR (obtained with the observed
H$\alpha$ or/and UV luminosities) and the obscured SFR (obtained with
the IR emission) requires the existence of a (simple) correlation
(close to pure proportionality) between the extincted correction for
the UV/optical SFR tracers and the emission in the IR. The right panel
of Figure~\ref{tir_halpha} explores this correlation for the TIR
luminosity, comparing it with the H$\alpha$ luminosity extinguished by
dust ($L^\mathrm{abs}(\mathrm{H}\alpha)$, calculated by subtracting
the observed luminosity from the extinction corrected luminosity) for
the HII regions in M81 and M51. The best linear fit to the data is:

\begin{displaymath}
\log\{L(8-1000)\}=\\
\end{displaymath}
\begin{equation}
(+8.9\pm1.0)+(+0.826\pm0.025)\times \log\{L^\mathrm{abs}(H\alpha)\}
\end{equation}

The scatter of the data around this relationship is 30\%. The scatter
of the points for the M81 data is larger than for the M51 data (45\%
and 20\%, respectively). Most of this scatter (and the scatter
observed in all the panels in Figures~\ref{tir_halpha} through
\ref{8_halpha}) is due to the larger uncertainties in the extinctions derived
for the HII regions in M81 in comparison with the
Pa$\alpha$-to-H$\alpha$ derived extinctions for M51. The scatter might
also be caused by spatial changes in the extinction properties within
an individual HII region (i.e., the actual net extinction and the
attenuation law). If the central parts of the HII regions had larger
extinctions than the outskirts, our values of
$L^\mathrm{corr}(\mathrm{H}\alpha)$ would be overestimated. To test
this effect, we measured fluxes for the same HII regions but with
smaller apertures. The correlation was slightly better in this case,
with a scatter around the linear relationship of 25\%.

The different trends shown by the M81 and M51 data in the inset of the
left panel of Figure~\ref{tir_halpha} disappear in this plot, although
the scatter of points is slightly larger, mostly because of the M81
points (and especially the ones with extinctions estimated with radio
data). In addition, the slope of the correlation differs significantly
from unity (0.826$\pm$0.025), the value that is expected if the two
quantities are closely correlated. This behavior suggests that the
heating of the colder dust, which dominates the TIR luminosity, is not
directly caused (exclusively) by the Lyman photons. The right panel of
Figure~\ref{tir_halpha} also shows the luminosity relation with slope
unity and a ratio between the H$\alpha$ and TIR luminosities given by
the SFR calibrations found in \citet{1998ARA&A..36..189K}. Note that
Kennicutt's calibration was built for entire starburst galaxies
assuming continuous star formation for the last 10--100~Myr. Despite
this, the relation fits the data points for the M81 HII regions very
well, but it departs significantly from the brightest M51 points. This
effect indicates differences in the attenuation properties (for
example, in the attenuation law) of the mildly extincted M81 regions
in comparison with the heavily attenuated M51 central zone.

\subsection{The rest-frame 24~\mic\, luminosity as a SFR tracer}
\label{lmipssfr}

\begin{figure*}
\begin{center}
\includegraphics[angle=-90,width=7.5cm]{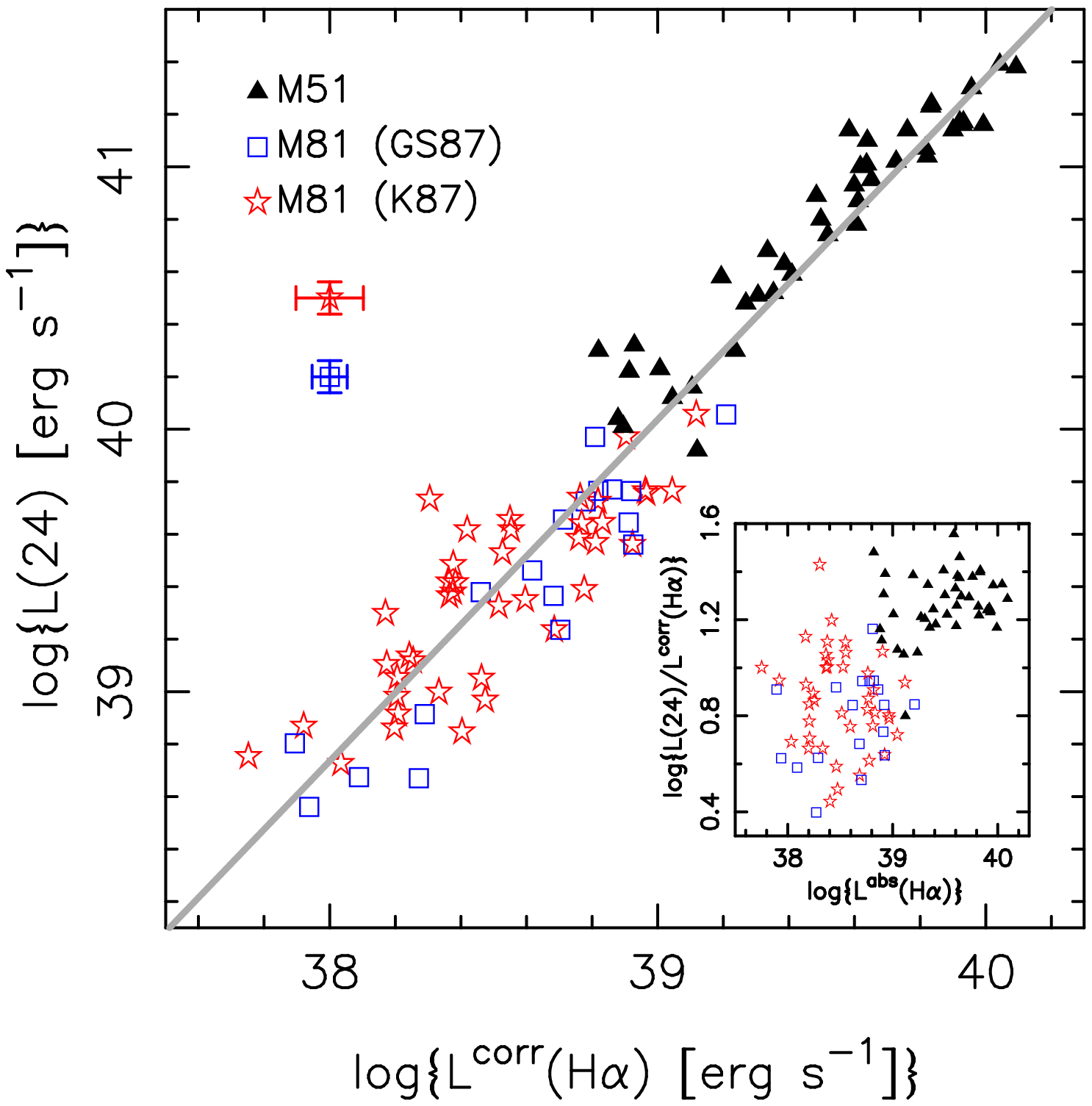}
\hspace{0.2cm}
\includegraphics[angle=-90,width=7.5cm]{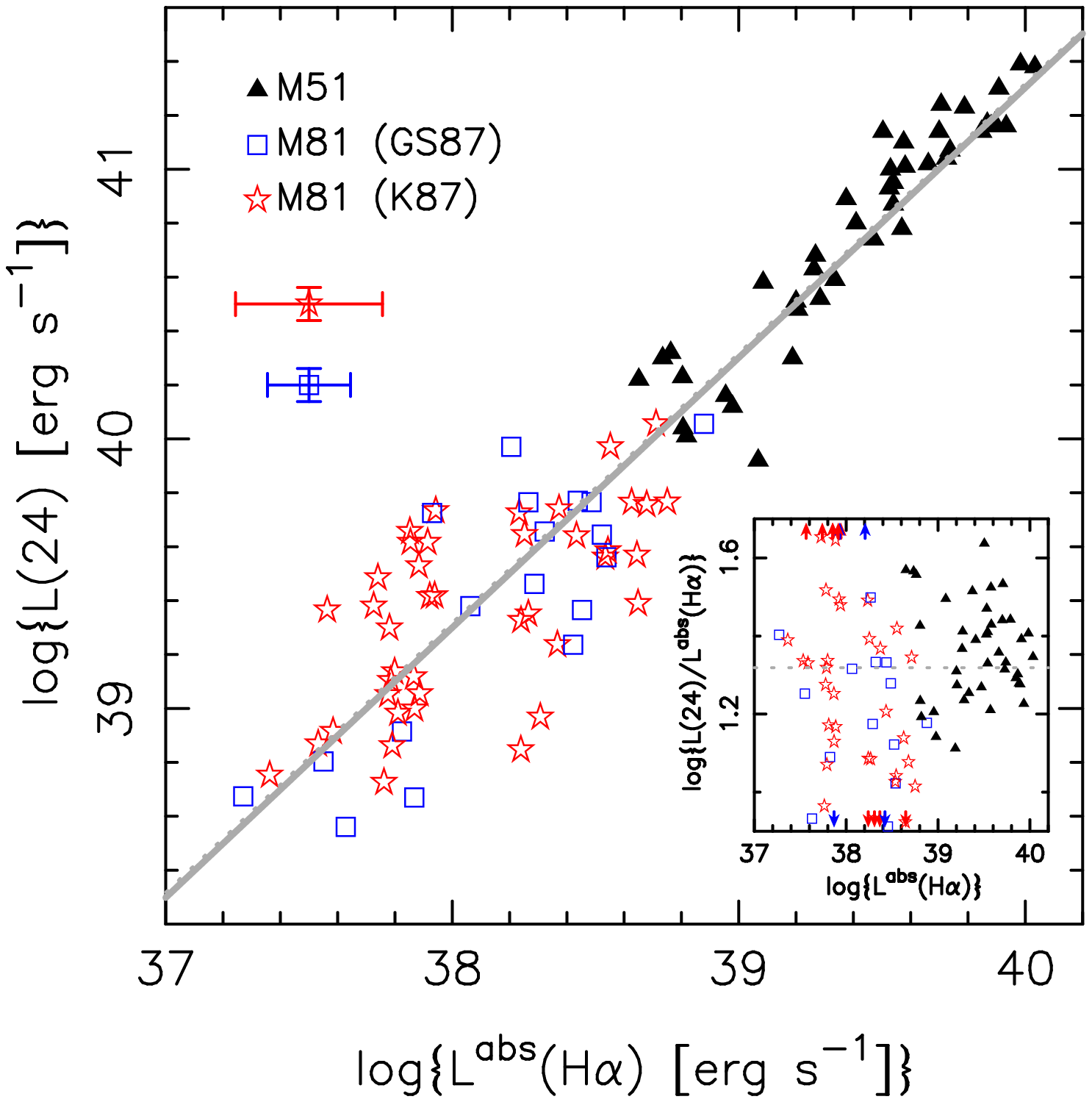}
\figcaption{\label{24_halpha}On the left, comparison between the 
extinction corrected H$\alpha$ and the 24~\mic\, luminosities for
selected HII regions in M81 (with different colors and symbols
referring to the distinct types of regions defined in the right panel
of Figure~\ref{reg_160}, extracted from \citealt{1987ApJ...317...82G}
-GS87- and \citealt{1987ApJ...319...61K} -K87-) and M51 (black
triangles; data extracted from \citealt{2005ApJ...633..871C}). On the
right, the 24~\mic\, luminosity of the same sample of HII regions is
compared with the H$\alpha$ luminosity absorbed by the dust.  For both
panels, we show the typical uncertainties linked to each catalog of
HII regions in M81. The best linear fit is plotted with a continuous
gray line. The insets show the same plots but using a luminosity ratio
in the vertical axis to enhance the differences in the trends followed
by the datasets for each galaxy. The dotted line in the right panel
(and the inset) shows the luminosity relation for slope unity.}
\end{center}
\end{figure*}

\slugcomment{Please, plot this figure with the width of one column}
\placefigure{24_halpha}

For the nuclear region of M51, \citet{2005ApJ...633..871C} found that
the 24~\mic\, emission correlates tightly (even better than the TIR
luminosity) with the extinction corrected Pa$\alpha$ luminosity,
suggesting that the warm dust responsible for the 24~\mic\, emission
is heated by the ionizing photons. Indeed, it seems reasonable that
the hottest dust in a galaxy could be more efficiently heated by the
very energetic photons arising from the hottest (and youngest) stars
than by redder photons from older stars. In contrast, the colder dust
could be heated not only by the ionizing flux, but also by the
emission arising from cooler (and older) stars. In such case, the
luminosities in the MIR would be a better tracer of the most recent
star formation than the TIR luminosity, which is dominated by the
emission of colder dust (see Equations~\ref{3mips_tir} and
\ref{3mips_tir_nodiffuse}). This is supported by the fact 
that the morphologies of M81 and M51 at 24~\mic\, resemble those of
the H$\alpha$ emission (see \citealt{2004ApJS..154..215G} and
\citealt{2005ApJ...633..871C}; see also  
\citealt{2004ApJS..154..259H} for M33, and 
\citealt{2004ApJS..154..253H} for NGC300), which closely traces the 
location of the ionizing stars (in the HII regions). In contrast, the
morphologies in the MIR are not as similar to the UV images, given
that the UV flux is largely affected by extinction and the emission is
normally linked to the star formation on a longer timescale.

We explore the possibility of using the 24~\mic\, emission alone as a
SFR tracer in Figure~\ref{24_halpha}. The left panel shows the
correlation between $L(24)$ and $L^\mathrm{corr}(\mathrm{H}\alpha)$
for the combined sample of HII regions in M81 and M51. The best fit to
the data is:

\begin{displaymath}
\log\{L(24)\}=\\
\end{displaymath}
\begin{equation}
(-10.7\pm1.4)+(+1.302\pm0.036)\times \log\{L^\mathrm{corr}(\mathrm{H}\alpha)\}
\end{equation}

The global scatter around this relationship is 40\%. As seen in the
left panel of Figure~\ref{tir_halpha}, the data points for M81 and M51
seem to fall on two different linear relations with very similar
slopes (very close to unity) but offset (see in the inset in the
left panel of Figure~\ref{24_halpha}). This effect is consistent with
the different $L(24)/SFR$ ratios found by
\citet{2005ApJ...633..871C} for the HII regions in M51 and for entire
galaxies, including starbursts and ultraluminous infrared
galaxies. While the galaxies with very intense star formation show
larger $L(24)/SFR$ ratios than the central region of M51, the HII
regions in M81 show smaller ratios (on average, a $L(24)/SFR$ ratio a
factor of three lower than the central region of M51). As pointed out
by \citet{2005ApJ...633..871C}, $L(24)$ alone does not seem to be a
universal estimator of the total SFR of a galaxy or HII
region. However, \citet{almudena06} have derived an empirical relation
that provides an accurate fit to the SFR (obtained from Pa$\alpha$
data) to 24~\mic\, luminosity for the dusty M51 HII regions upward in
the luminosity through $L(8-1000)\sim10^{12}\,L_\sun$.

In the right panel of Figure~\ref{24_halpha}, we show the alternative
calibration of the 24~\mic\, luminosity as a tracer of the amount of
H$\alpha$ photons extinguished by dust, i.e., not of the total star
formation, but only of the star formation which is undetectable by the
observed H$\alpha$ emission due to dust extinction. The best fit to
the data is:

\begin{displaymath}
\log\{L(24)\}=\\
\end{displaymath}
\begin{equation}
\label{24_calib}
(+1.2\pm1.3)+(+1.002\pm0.034)\times \log\{L^\mathrm{abs}(\mathrm{H}\alpha)\}
\end{equation}

\begin{figure*}
\begin{center}
\includegraphics[angle=-90,width=7.5cm]{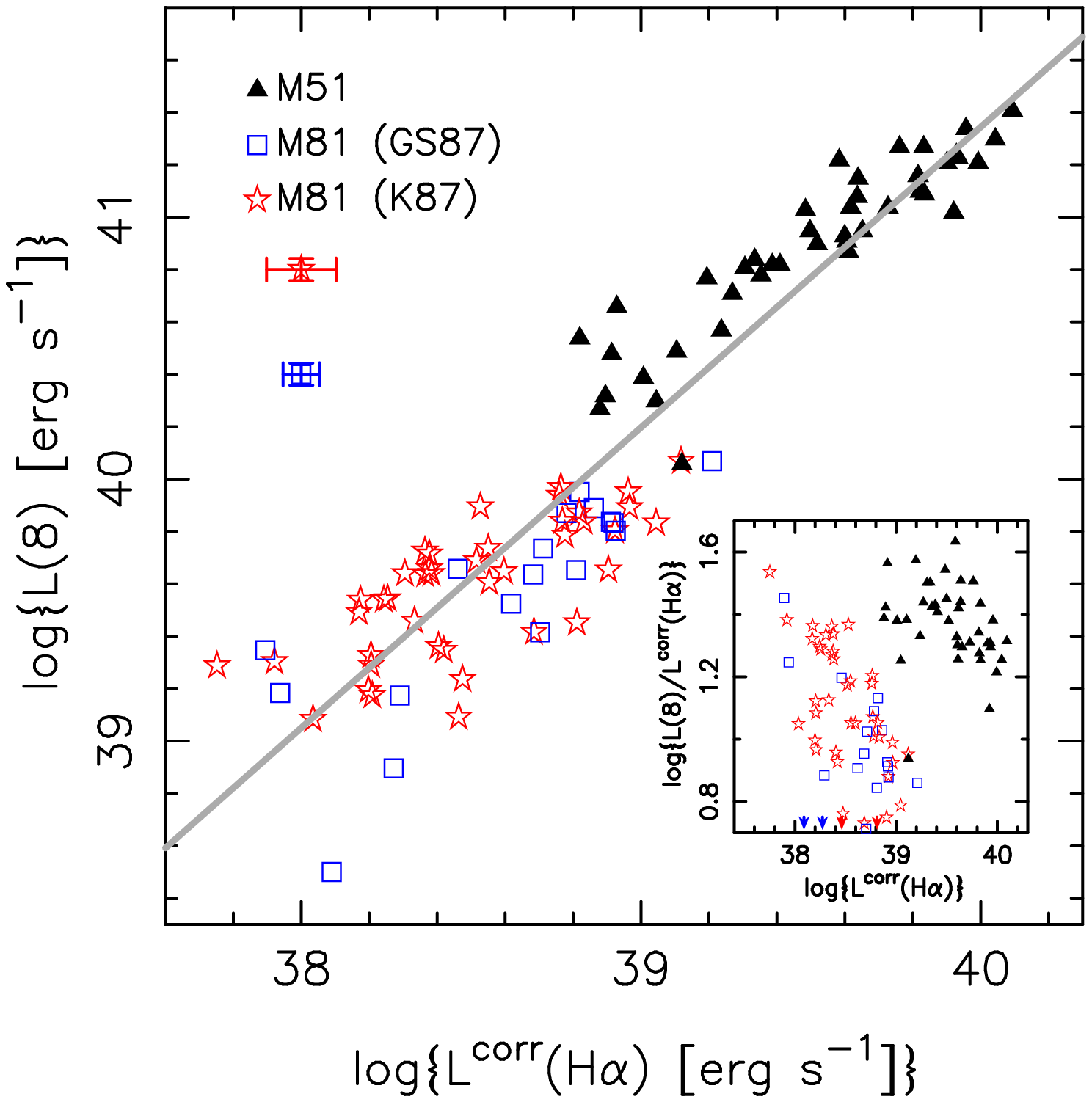}
\hspace{0.2cm}
\includegraphics[angle=-90,width=7.5cm]{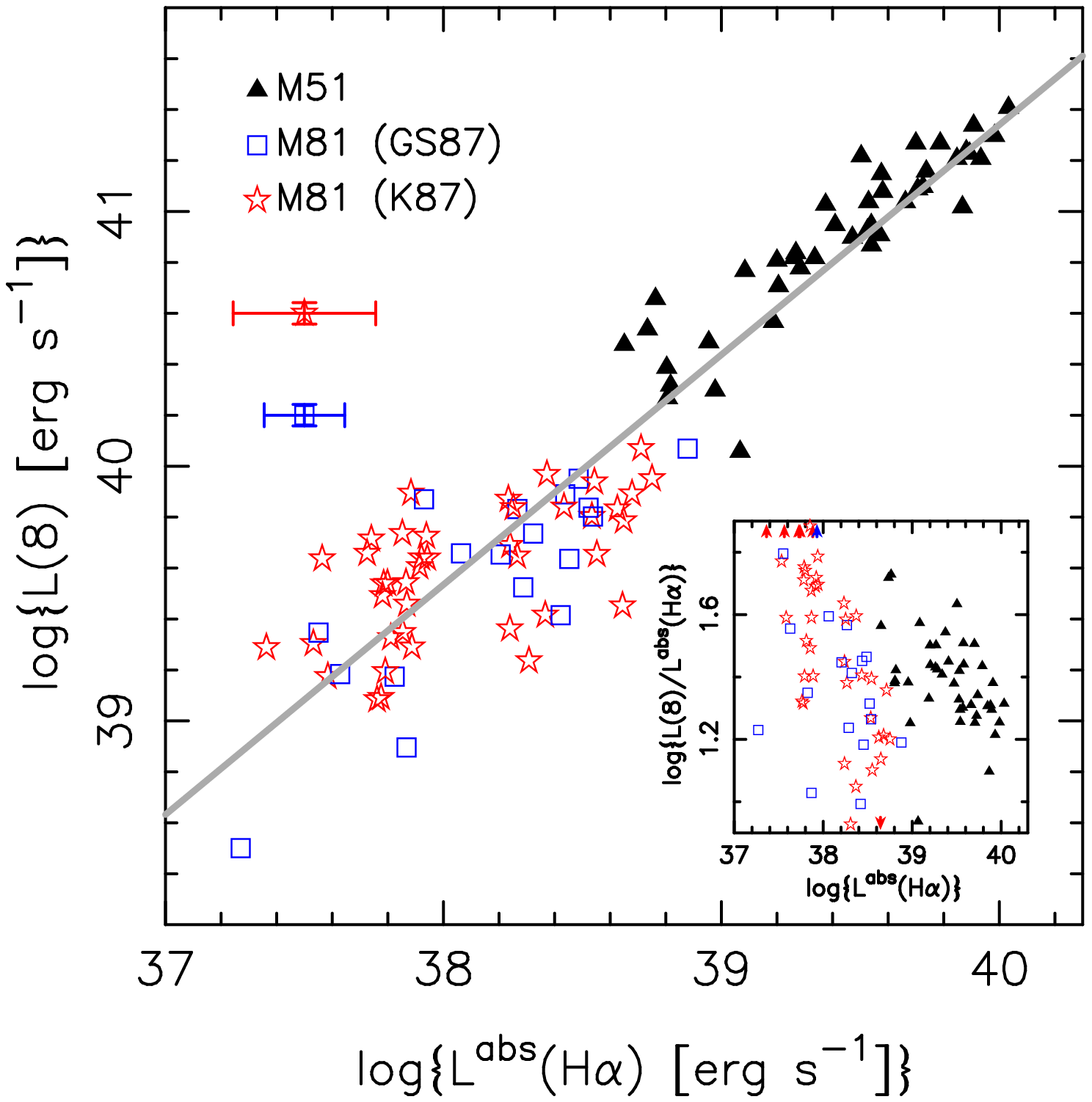}
\figcaption{\label{8_halpha}On the left, comparison between the 
extinction corrected H$\alpha$ and the 8~\mic\, luminosities for
selected HII regions in M81 (with different colors and symbols
referring to the distinct types of regions defined in the right panel
of Figure~\ref{reg_160}, extracted from \citealt{1987ApJ...317...82G}
-GS87- and \citealt{1987ApJ...319...61K} -K87-) and M51 (black
triangles; data extracted from \citealt{2005ApJ...633..871C}). On the
right, the 8~\mic\, luminosity of the same sample of HII regions is
compared with the H$\alpha$ luminosity absorbed by the dust. For both
panels, we show the typical uncertainties linked to each catalog of
HII regions in M81. The best linear fit is plotted with a continuous
gray line. The insets show the same plots but using a luminosity ratio
in the vertical axis to enhance the differences in the trends followed
by the datasets for each galaxy.}
\end{center}
\end{figure*}

\slugcomment{Please, plot this figure with the width of one column}
\placefigure{8_halpha}

The global scatter around this relationship is 35\%. The most deviant
points correspond again to the HII regions in M81 with a radio-based
extinction. The slope of the fit is consistent with a value of unity,
which favors a direct link between the Lyman photons (or the emission
arising from the stars that dominate the production of Lyman photons)
and the heating of the hottest dust in M81 and M51. Note that the SFR
calculated in this way should be added to the SFR estimated from the
observed emission-line to obtain the total SFR.

\subsection{The 8~\mic\, luminosity as a SFR tracer}

Figure~\ref{8_halpha} shows the comparison of the 8~\mic\, luminosity
(corrected for the stellar light contamination) with the extinction
corrected H$\alpha$ luminosity and the H$\alpha$ luminosity
extinguished by dust. To be able to compare with our data, the
8~\mic\, emission for the regions in M51 has been corrected for the
effect of the calibration of extended sources in IRAC images (a
correction of 0.133~dex). The best linear fits to the data are:

\begin{displaymath}
\log\{L(8)\}=\\
\end{displaymath}
\begin{equation}
(-4.5\pm1.6)+(+1.147\pm0.042)\times \log\{L^\mathrm{corr}(\mathrm{H}\alpha)\}
\end{equation}

\begin{displaymath}
\log\{L(8)\}=\\
\end{displaymath}
\begin{equation}
(+5.2\pm1.2)+(+0.902\pm0.031)\times \log\{L^\mathrm{abs}(\mathrm{H}\alpha)\}
\end{equation}

The scatter around these relations is larger than 60\% in both
cases. Comparing with Figures~\ref{tir_halpha} and \ref{24_halpha},
Figure~\ref{8_halpha} shows that the 8~\mic\, emission is a poorer
estimator of the total or extincted SFR than the 24~\mic\, or TIR
emissions. Within each galaxy, the data seem to follow a linear
relation (with a slope that differs significantly from unity), but
there are strong differences between the two galaxies. Three factors
can be responsible for these differences: 1) an important residual
contamination in the 8~\mic\, fluxes from stellar emission, which
might depend on the size of the apertures used and the properties of
the stellar population enclosed; 2) galaxy-to-galaxy variations in the
metallicity and aromatic properties and spectrum; and 3) a
heterogeneous nature in the agents responsible for the excitation of
the aromatic molecules that dominate the 8~\mic\, emission
\citep[see also][and references therein]{2005ApJ...633..871C}.

All the correlations found in this Section are summarized in
Table~\ref{sfr_corrs}.

\subsection{Summary of the IR SFR tracers}

All the results for M81 and M51 described in the previous subsections
support the idea that the heating of the dust emitting at 24~\mic\,
(the dust at $T=40-50$~K) is dominated by the light arising from the
same stars that produce most of the ionizing flux, which is directly
correlated with the emission-line fluxes. In addition, the spectrum
around 24~\mic\, shows a remarkable absence of any bright emission or
absorption feature (such as aromatic molecules) for a wide variety of galaxies
\citep[see, e.g.,][]{2000A&A...358..481S,2004ApJS..154..178A,
2004ApJS..154..184S, 2004ApJS..154..199S,2004ApJS..154..211H}. This
means that the 24~\mic\, band is adequate to directly measure the
continuous emission from warm dust (heated by ionizing photons)
without line contaminants, in contrast to the significant
contamination of other parts of the IR spectrum by aromatic emissions
(e.g., 8 or 12~\mic) or/and emission from dust heated by sources not
directly linked with the recent star formation (e.g., 160~\mic). All
these reasons support the idea that the 24~\mic\, emission alone is a
good estimator of the obscured SFR for HII regions in a wide range of
extinctions (from $A(V)\sim0$ to $A(V)\sim4$) and SFR (from $10^{-4}$
to 0.3~$\mathcal{M}_\sun\,\mathrm{yr}^{-1}$). In the same sense, the
8~\mic\, emission is a poorer estimator of the SFR. But the IR
luminosities alone are not directly proportional to the total SFR for
HII regions in different galaxies. Regions with higher extinctions
present larger SFRs (see \citealt{2001AJ....122..288H},
\citealt{2001ApJ...558...72S}, and \citealt{2003ApJ...591..827P}) and 
higher IR-to-H$\alpha$ ratios that increase non-linearly with
$L^\mathrm{corr}(H\alpha)$. There are two reasons for this behavior:
1) the absorption of photons by dust does not scale linearly with the
radiation field seen by the dust; 2) as we increase the extinction,
larger zones of the HII regions become optically thick and the
emission-line observations are not able to recover all the emitted
flux. More data for other galaxies (spanning more extinction, SFRs,
metallicities, and stellar population properties) should be added to
this study to prove the universality of the correlations given in the
previous subsections.

\subsection{Calibration of the UV luminosity as a SFR tracer compatible with H$\alpha$ and the IR}

\begin{figure*}
\begin{center}
\includegraphics[angle=-90,width=4.5cm]{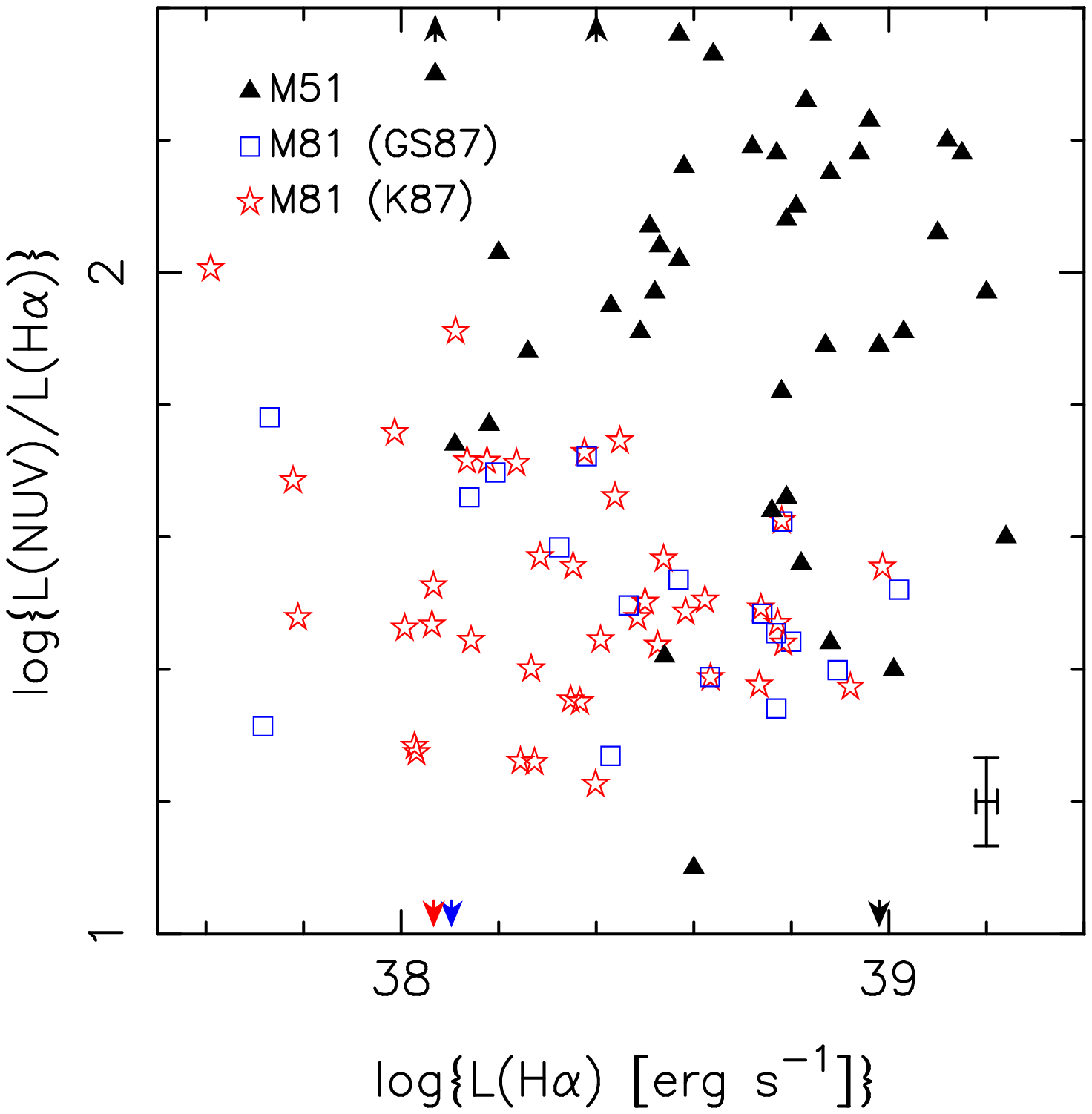}
\hspace{0.2cm}
\includegraphics[angle=-90,width=4.5cm]{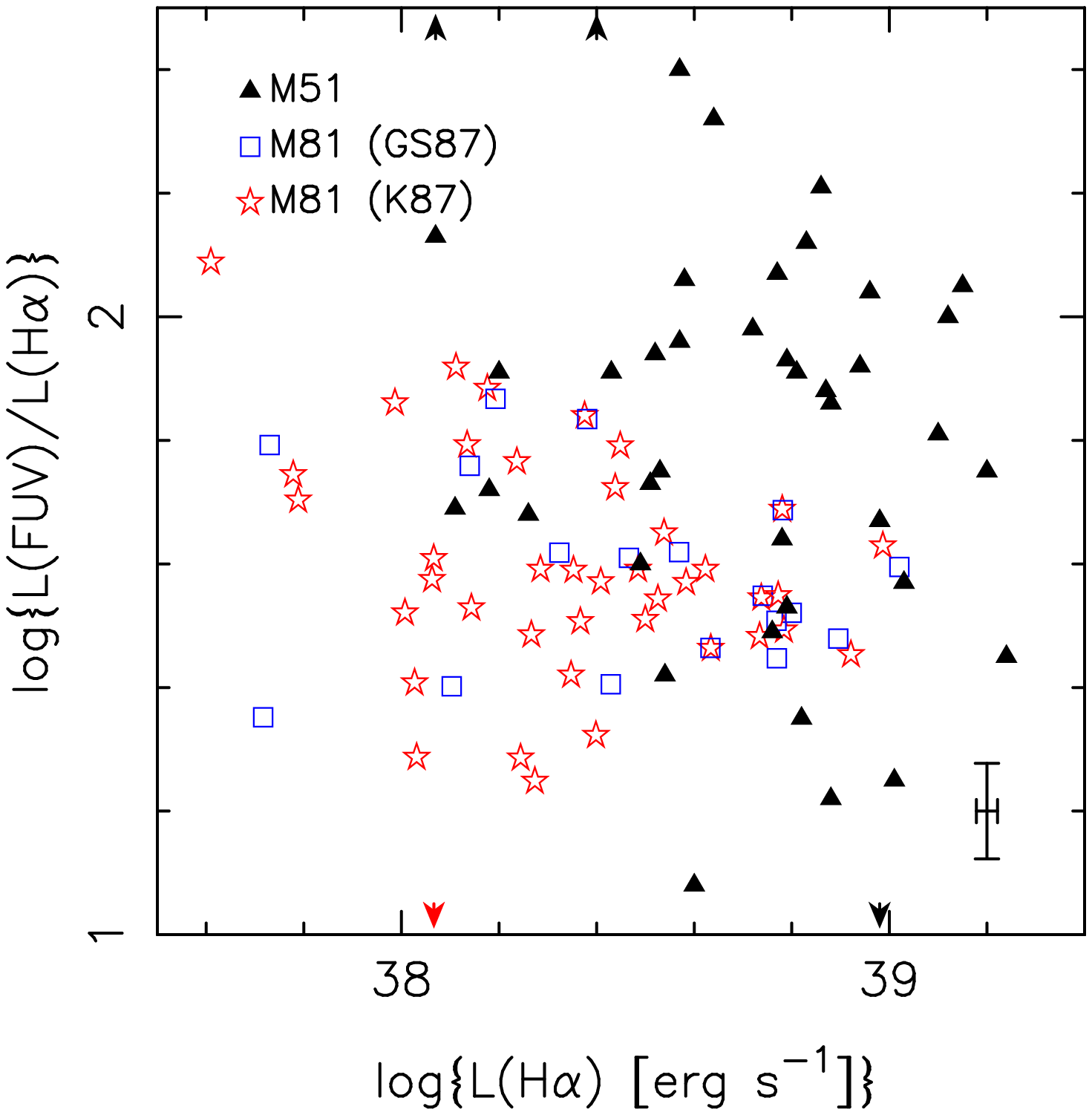}
\hspace{0.2cm}
\includegraphics[angle=-90,width=4.5cm]{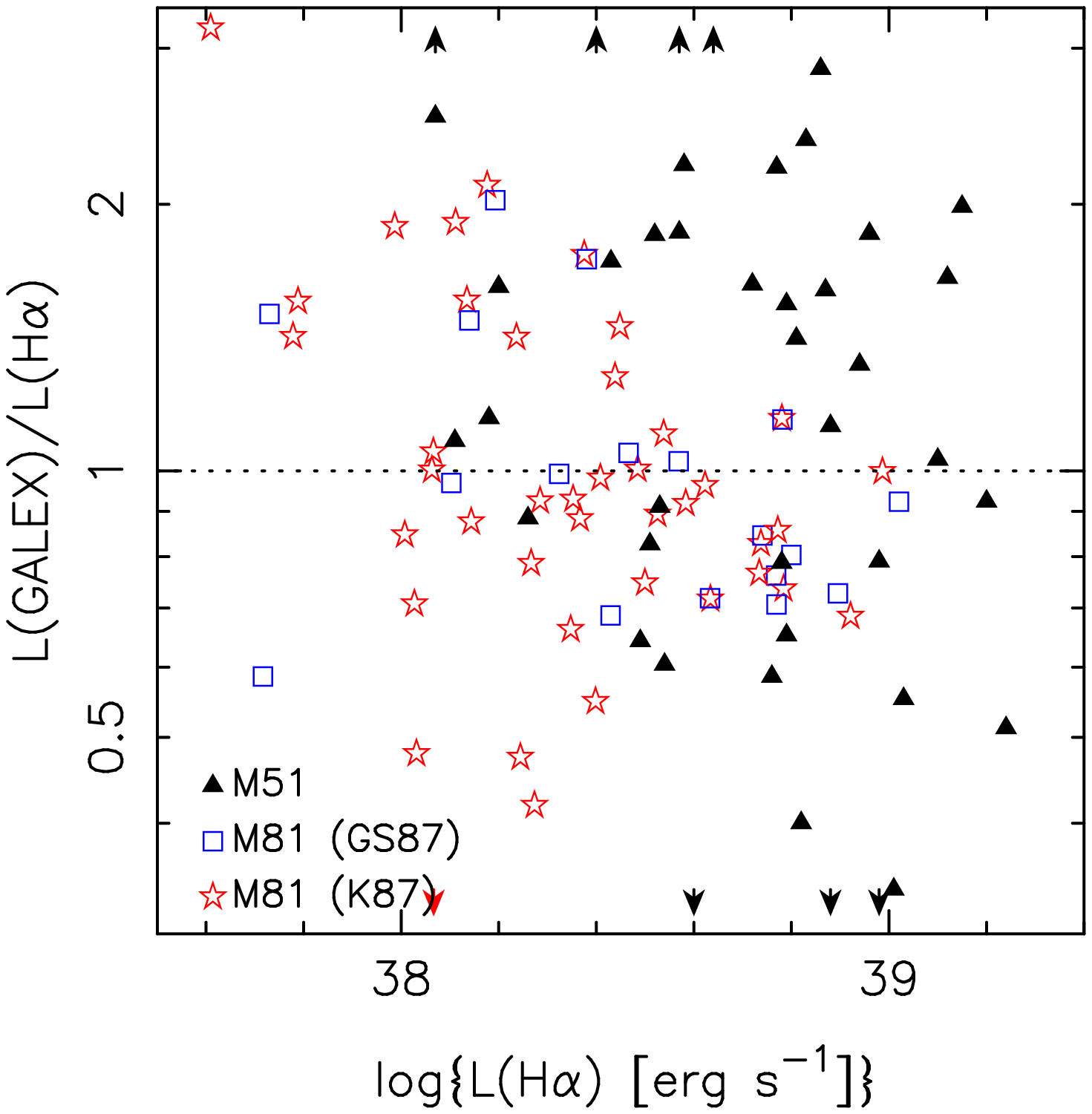}
\figcaption{\label{uv_halpha}In the first two panels, we show the 
comparison between the observed luminosities in the two GALEX UV bands
and the observed H$\alpha$ emission. In the right panel, we compare
the observed H$\alpha$ emission with a combined UV luminosity obtained
with the two GALEX bands, $L(GALEX)$ (defined in
Equation~\ref{galex_ha}), built to match the observed H$\alpha$
emission. In this panel, the expected luminosity ratio of unity (if
$L(GALEX)$ and $L(\mathrm{H}\alpha)$ match perfectly) is also shown.}
\end{center}
\end{figure*}

\slugcomment{Please, plot this figure with the width of one column}
\placefigure{uv_halpha}

In this Subsection, we compare the UV-derived SFRs with the H$\alpha$
observations and obtain a calibration of the UV data as a SFR tracer
compatible with the H$\alpha$ emission. This calibration will allow
the use of UV, H$\alpha$, and IR observations to analyze in detail the
SFRs and extinction properties of HII regions.

The relationship between the luminosity in the FUV or NUV and the SFR
is well known, given that the UV emission arises mainly from hot
massive stars. However, the emission at these wavelengths is strongly
affected by dust attenuation. This attenuation has been commonly
estimated using the slope in the UV part of the spectrum.  The UV
slope was calibrated against the actual extinction in the UV, measured
with the ratio between the IR and UV emissions, for a sample of
starburst galaxies by \citet{1999ApJ...521...64M}. This calibration
has been extensively used in the literature, but several studies have
shown that the relationship found for starbursts cannot be applied to
all galaxies \citep[see, e.g.,][]{2002ApJ...577..150B,
2004MNRAS.349..769K,2005ApJ...619L..51B}. Star-forming regions in M81
\citep{2004ApJS..154..215G} and M51 \citep{2005ApJ...633..871C} 
have also been shown to depart significantly from the starburst
relationship. All these works have shown that the relationship between
the UV slope and the extinction depends on factors such as the age of
the stellar population, the metallicity and the attenuation law (which
depends on the dust properties and the dust-stars geometry).  The
difficulties in extinction correction prevent the use of the UV
emission alone as an accurate SFR tracer.

Another possibility is to use the UV emission to estimate the SFR not
obscured by dust, and to add this component to the obscured SFR
estimated with the IR emission. This method is similar to the one used
in \citet{2003ApJ...586..794B}. It is also analogous to the procedure
we followed in the previous subsections, where we showed the good
correlation between the 24~\mic\, luminosity and the SFR that is
obscured by dust and cannot be detected with an emission-line such as
H$\alpha$ or Pa$\alpha$. The narrow-band data for those lines are very
expensive to obtain in terms of observing time. Therefore, it would be
convenient to have an estimate of the SFR not hidden by dust using
other observations which were relatively easier to obtain, such as the
GALEX UV data. The goal is to translate the broad-band UV emission to
an integrated emission in the UV, and that to a non-obscured SFR
consistent with the ionized hydrogen emission. In principle, the
H$\alpha$ emission traces a stellar population with a shorter
timescale than the UV (about 10~Myr for H$\alpha$ and 100~Myr for the
UV), given that the emission-line is linked to the hottest (and
youngest) stars dominating the production of ionizing photons. The
H$\alpha$/UV ratio should be more and more independent of the star
formation history of the galaxy as we move to shorter wavelengths in
the UV \citep{2002A&A...383..801B,2002ApJ...577..150B}. The extinction
also plays an important role, since it affects the stellar and ionized
gas emission differently \citep[see,
e.g.,][]{2000ApJ...533..682C,2000ApJ...539..718C,2002ApJ...565..994B}.

In Figure~\ref{uv_halpha}, we plot the relationships between the
observed H$\alpha$ emission and the luminosities in the two GALEX UV
bands for the HII regions in M81 and the central part of M51. The left
panel shows that the NUV luminosity is not directly proportional to
the H$\alpha$ emission (a proportionality would show in the figure as
a cloud of points around an average value, but there is an offset
between the M51 and M81 points). The lack of correlation is related to
the fact that the NUV emission arises from cooler stars than the ones
that dominate the production of ionizing photons (i.e., the difference
in the timescales of the star formation traced by H$\alpha$ and the
NUV is relatively large). The different luminosity ratio found for M81
and M51 should also be related to the larger (extinction corrected)
SFRs observed for the regions in M51. The correlation between the FUV
and H$\alpha$ emission is considerably better (i.e., they are roughly
proportional), although there still seems to be an offset between the
M81 and M51 data. This panel shows that the FUV luminosity probes a
stellar population that is closer to that producing the Lyman
photons, as expected.

The first two panels of Figure~\ref{uv_halpha} show that there is not
a direct correlation between the broad-band UV emission and the
observed H$\alpha$ luminosity, since parameters such as the age of the
stellar population and the extinction affect both quantities in
differing amounts. In the last panel of Figure~\ref{uv_halpha}, we
explore the possibility of using the two GALEX luminosities to obtain
an estimation of the unobscured SFR consistent with the observed
H$\alpha$ emission. For this purpose, we obtained a relationship
between the observed H$\alpha$ luminosity and the emission in the two
UV bands using a singular value decomposition method. By using the two
UV fluxes, the relation is intended to cope with the two parameters
affecting the UV slope, the age and the extinction (for a fixed
metallicity), and to allow obtaining an estimate of the unobscured SFR
from UV data consistent with the observed H$\alpha$ flux. The result
is the following:

\begin{displaymath}
\log\{L(\mathrm{H}\alpha)\}=\\
\end{displaymath}
\begin{equation}
\label{galex_ha}
+1.437\times\log\{L(FUV)\}+(-0.478)\times\log\{L(NUV)\}\equiv \log\{L(\mathrm{GALEX})\}
\end{equation}

This empirical relationship allows estimating the unobscured SFR using
the GALEX data with a scatter of 40\%. More data for other galaxies
should be added to this study to prove the universality and
usefulness of such a correlation.

\subsection{The radial trend in the dust attenuation}

\begin{figure}
\begin{center}
\includegraphics[angle=-90,width=8.5cm]{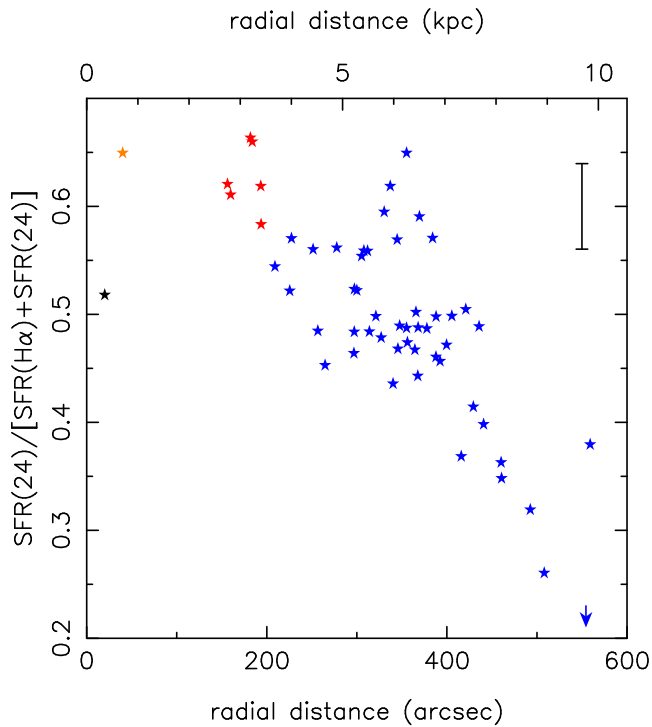}
\figcaption{\label{radial_grad}Radial dependence (de-projected) 
of the ratio between the obscured SFR estimated from the 24~\mic\,
luminosity (using the calibration in Equation~\ref{24_calib}) and the
total SFR calculated with the observed H$\alpha$ and 24~\mic\,
emissions. The different colors refer to the distinct types of regions
defined in Figure~\ref{reg_160}. The average surface density of the entire galaxy
has been plotted at $r=20\arcsec$, and the surface density for the
nuclear and the circumnuclear regions has been plotted at
$r=40\arcsec$ (slightly shifted from $r=0$ for clarity).}
\end{center}
\end{figure}

\slugcomment{Please, plot this figure with the width of one column}
\placefigure{radial_grad}

Figure~\ref{radial_grad} shows a radial plot of the ratio between the
obscured SFR (calculated with the 24~\mic\, luminosity, translated to
an equivalent absorbed H$\alpha$ luminosity with
Equation~\ref{24_calib}, and from that to a SFR using the calibration
in \citealt{1998ARA&A..36..189K}) and the total SFR (calculated as the
sum of the unobscured and obscured SFRs with the observed H$\alpha$
and 24~\mic\, emissions) for the regions defined in the 160RES image
set\footnote{We have used the 160RES regions because they span a wider
range of radial distances than the 24RES regions} and using the
LOCALBKG photometry. There is a clear gradient in the SFR ratio, i.e.,
there is a decrease in the extinction of the stellar light as we move
away from the center of the galaxy. This gradient is consistent with
the results we found in Section~\ref{results_1} about the radial trend
in two other quantities: the dust temperature and the dust mass
surface density. It is also consistent with results obtained for M81
and other galaxies in the literature \citep{2004A&A...424..465B,
2005A&A...444..109H,2005ApJ...632..227R,prescott06}.

Figure~\ref{radial_grad} also shows that for the entire M81 galaxy,
the IR just sees $\sim$50\% of the total SFR; the other 50\% is linked
with star formation that is not obscured by dust. The largest
percentages are observed for the regions dominated by very diffuse
emission, where about 65\% of the star formation is seen in the
IR. For the most remote regions in the disk, the IR alone can miss up
to $\sim$80\% of the star formation. The gradient in the SFR ratio
implies a radial change in the H$\alpha$ extinction from
$A(\mathrm{H}\alpha)\sim1.0$ to $A(\mathrm{H}\alpha)\sim0.2$ along the
disk of M81, the observed range for the 24RES regions used in the
previous subsections.

\input{tab4_2c}

\section{SUMMARY AND CONCLUSIONS}
\label{conclusions}

The recent star formation in the early-type spiral galaxy M81 has been
analyzed in two dimensions using imaging observations spanning the
electromagnetic spectrum from the far ultraviolet (UV) to the far
infrared (IR). The data have been compared to models of stellar, gas,
and dust emission, obtaining results for different sub-galactic
regions whose sizes range from those typical of individual HII regions
(0.1~kpc) to the sizes of dynamical structures (2~kpc).

The main results of this work are:

\begin{itemize}
\item[1.-]Using stellar population models, we confirm that the 
luminosities in the mid-IR bands (from 3 to 8~\mic) contain
substantial contributions from both dust and stellar light. These
contributions vary significantly depending on the spatial resolution
and environment, suggesting the existence of mid-and far-IR diffuse
dust emission not directly linked to the recent star formation.

\item[2.-] The IR emission of M81 can be modeled with three 
components: 1) a cold dust component of average temperature
$T_c=18\pm2$~K; 2) a warm component with $T_w=53\pm7$~K; and 3)
polycyclic aromatic molecules with different properties. While the
emission of the hotter dust (peaking at around 24~\mic) is very
concentrated in the HII regions, the emission of the colder dust
component (peaking near 160~\mic) and the aromatic molecules
(contributing importantly to the 8~\mic\ emission) show a more diffuse
distribution. This suggests the presence of an alternative heating
source different than the stars recently formed in the HII regions
(older stars in the disk), or the importance of the transfer of energy
(e.g., the ionizing flux) from the HII regions to the interstellar
medium (i.e., the escape of photons from the HII regions that
interact with dust in distant zones of the disk).

\item[3.-] A negative radial gradient is observed in the dust 
temperature of the cold dust component, the dust mass surface density
(similar to the gradient of the molecular gas surface density found by
\citealt{1995AJ....110.1115D}), and the comparison between the star
formation rate (SFR) indicators affected by dust extinction ($UV$ and
H$\alpha$ emission) and the IR emission \citep[see
also][]{2005ApJ...632..227R,prescott06}.

\item[4.-] For the entire M81 galaxy, about 50\% of the star 
formation is obscured by dust and can be seen in the IR, and the other
half can be traced directly with the observed H$\alpha$ emission. The
percentage of obscured star formation ranges from 60\% in the inner
regions of the galaxy to 30\% in the most distant zones.

\item[5.-] Based on our dust emission models, we have built 
several empirical relations between the monochromatic fluxes in the IR
and the total IR emission [integrated between 8 and 1000~\mic,
$L(8-1000)$], useful for sub-galactic regions in spiral galaxies with
$10^7\lesssim L(8-1000)\lesssim 10^{8}$~L$_\sun$. These correlations
differ significantly from the ones found in the literature for entire
IR-bright galaxies, mainly because of the presence of a colder dust
component.

\item[6.-] We have compared five different SFR
estimators for individual HII regions in M81 and M51: the UV, the
H$\alpha$ emission, and three estimators based on \spitzer\, IR
data. Several empirical relations have been built to estimate the
total SFR of HII regions from IR data alone, or from a combination of
IR fluxes (which account for the obscured star formation) and
UV/optical data (directly measuring the unobscured star
formation). The typical scatter of data around these relations is
30-50\%, with possible important additional variations of the
correlations from galaxy to galaxy.

\item[7.-]Among the SFR estimators in the IR, the 
24~\mic\, emission has the best correlation with the H$\alpha$
luminosity extinguished by dust. The correlation is worse for the
total IR luminosity, which suggests that the colder dust dominating
the total IR luminosity is (also) heated by redder photons. The
8~\mic\, luminosity is not directly correlated with the ionized gas
emission, making it in an unreliable SFR estimator. These results are
consistent with those found in
\citet{2005ApJ...633..871C} and
\citet{kennicutt06}.

\item[8.-]Important variations from galaxy to galaxy are
found when estimating the total SFR with the 24~\mic\, or the total IR
emission alone. We suggest the combination of the H$\alpha$ (or the
UV) and an IR luminosity (especially the rest-frame 24~\mic\,
emission, calibrated in \citealt{kennicutt06} and in this paper with
the M81 and M51 data) to obtain the most reliable estimates of the
total SFRs for HII regions (and, by extrapolation, for
galaxies). Those emissions probe the unobscured and obscured star
formation, respectively.

\item[9.-]Additional data for other galaxies (spanning more
extinction, SFRs, metallicities, and stellar population properties)
should be added to this study to prove the universality of the
correlations found in this paper.

\end{itemize}

\acknowledgments
We thank an anonymous referee for her/his useful comments. Support for
this work was provided by NASA through Contract no.  1255094 issued by
JPL/Caltech. This work is part of SINGS, the Spitzer Infrared Nearby
Galaxies Survey, one of the Spitzer Space Telescope Legacy Science
Programs, and was supported by the JPL, Caltech, contract 1224667.
This work is based in part on observations made with the {\it Spitzer}
Space Telescope, which is operated by the Jet Propulsion Laboratory,
Caltech under NASA contract 1407. GALEX is a NASA Small Explorer
launched in 2003 April. We gratefully acknowledge NASA's support for
construction, operation, and scientific analysis of the GALEX
mission. This research has made use of the NASA/IPAC Extragalactic
Database (NED) which is operated by the Jet Propulsion Laboratory,
California Institute of Technology, under contract with the National
Aeronautics and Space Administration. P.G.~P.-G. and A.~G. de P.  also
wish to acknowledge support from the Spanish Programa Nacional de
Astronom\'{\i}a y Astrof\'{\i}sica under grant AYA 2004-01676.

\bibliographystyle{apj}
\bibliography{referencias}

\end{document}

%% file: tab1_2c.tex
\placetable{photometry_160RES}
\begin{deluxetable*}{lcccccccccc}
\scriptsize
\tablecaption{\label{photometry_160RES}Positions and photometry (GLOBALBKG and LOCALBKG cases) for the regions selected at 160~$\mu$m resolution.}
\tablehead{\colhead{Field} & \colhead{BKG} & \colhead{M81} & \colhead{Reg02} & \colhead{Reg03} & \colhead{Reg04} & \colhead{Reg05} & \colhead{Reg06} & \colhead{Reg07} & \colhead{Reg08} & \colhead{Reg09}}
\startdata
RA (J2000)              &   & 09:55:32.2 & 09:55:32.2 & 09:55:32.2 & 09:55:16.7 & 09:55:05.0 & 09:55:35.1 & 09:55:53.3 & 09:55:35.7 & 09:55:26.4\\
DEC (J2000)             &   & 69:03:59.0 & 69:03:59.0 & 69:03:59.0 & 69:06:12.4 & 69:05:22.9 & 69:06:24.1 & 69:01:47.0 & 69:07:29.6 & 69:08:08.9\\
radius                  &   & 680.0 & 64.0 & 104.0 & 40.9 & 50.8 & 51.3 & 56.0 & 31.9 & 31.6\\
$\log{[L(H\alpha)]}$    & G & 40.78 & 39.67 & 39.40 & 38.79 & 38.88 & 39.10 & 38.86 & 39.10 & 39.24\\
                        & L & 40.69 & 39.53 & 39.18 & 38.40 & 38.42 & 38.82 & 38.54 & 39.06 & 39.09\\
$\log{[L(3.6)]}$        & G & 43.18 & 42.63 & 42.38 & 41.63 & 41.62 & 41.75 & 41.81 & 41.14 & 41.10\\
                        & L &  \nodata &  \nodata &  \nodata &  \nodata &  \nodata &  \nodata &  \nodata &  \nodata &  \nodata\\
$\log{[L(4.5)]}$        & G & 42.84 & 42.28 & 42.03 & 41.27 & 41.25 & 41.40 & 41.49 & 40.81 & 40.77\\
                        & L &  \nodata &  \nodata &  \nodata &  \nodata &  \nodata &  \nodata &  \nodata &  \nodata &  \nodata\\
$\log{[L(5.8)]}$        & G & 42.63 & 41.98 & 41.76 & 41.05 & 41.06 & 41.21 & 41.21 & 40.73 & 40.72\\
                        & L & 41.42 &  \nodata &  \nodata &  \nodata & 39.30 & 40.03 &  \nodata & 40.32 & 40.22\\
$\log{[L(8.0)]}$        & G & 42.61 & 41.76 & 41.58 & 40.93 & 40.99 & 41.18 & 41.11 & 40.88 & 40.88\\
                        & L & 42.33 & 40.65 & 40.79 & 40.34 & 40.42 & 40.78 & 40.58 & 41.05 & 40.70\\
$\log{[L(24)]}$         & G & 41.98 & 41.25 & 40.90 & 40.20 & 40.29 & 40.49 & 40.38 & 40.28 & 40.35\\
                        & L & 41.92 & 41.13 & 40.68 & 39.90 & 39.96 & 40.31 & 40.20 & 40.27 & 40.28\\
$\log{[L(70)]}$         & G & 42.67 & 41.83 & 41.62 & 40.98 & 41.00 & 41.23 & 41.15 & 40.98 & 41.05\\
                        & L & 42.68 & 41.79 & 41.52 & 40.79 & 40.72 & 41.11 & 41.08 & 40.98 & 41.00\\
$\log{[L(160)]}$        & G & 42.99 & 41.78 & 41.78 & 41.27 & 41.37 & 41.55 & 41.46 & 41.30 & 41.31\\
                        & L & 42.99 & 41.67 & 41.62 & 41.05 & 41.11 & 41.38 & 41.36 & 41.29 & 41.22\\
\enddata
\tablecomments{Table 1 is available in its entirety via the link to the 
machine-readable version above. Units of right ascension are hours, 
minutes, and seconds, and units of declination are degrees, 
arcminutes, and arcseconds. The radius of each region is given in 
arcsec. All luminosities refer to observed values (no internal or 
Galactic extinction corrections have been applied) and the units are 
erg~s$^{-1}$. The first row for each wavelength refers to the 
photometry measured with a GLOBALBKG estimation (second column set to 
the value G), and the second to the LOCALBKG case (second column set to the value L).}
\end{deluxetable*}

%% file: tab2_2c.tex
\placetable{photometry_24RES}
\begin{deluxetable*}{lcccccccccc}
\tablecaption{\label{photometry_24RES}Positions and photometry for the regions selected at 24~$\mu$m resolution.}
\tablehead{ \colhead{Field} & \colhead{M81} & \colhead{kauf240} & \colhead{kauf104} & \colhead{kauf101} & \colhead{Munch1} & \colhead{kauf102} & \colhead{kauf198} & \colhead{No17} & \colhead{kauf197}}
\startdata
RA (J2000)               & 09:55:33.2 & 09:56:27.0 & 09:56:20.6 & 09:56:19.9 & 09:56:18.9 & 09:56:18.3 & 09:56:17.2 & 09:56:17.2 & 09:56:16.6\\
DEC (J2000)              & 69:03:55.1 & 69:04:24.0 & 69:01:12.1 & 69:04:25.3 & 68:49:40.9 & 69:03:34.9 & 69:06:07.4 & 69:06:07.4 & 69:05:36.2\\
radius                   &   680.0  &    8.4  &    8.4  &    8.4  &    8.4  &    8.4  &    8.4  &    8.4  &    8.4 \\
$A(V)$                   & \nodata  &   1.05   &   1.31   &   1.01   &   0.37   &   1.91   &   0.42   &   1.60   &   0.67  \\
$\log{[L(FUV)]}$         &  42.74   &  39.49   &  39.67   &  40.00   &  39.51   &  39.98   &  40.22   &  40.22   &  39.88  \\
$\log{[L(NUV)]}$         &  42.82   &  39.25   &  39.57   &  39.93   &  39.06   &  39.83   &  40.08   &  40.08   &  39.84  \\
$\log{[L(H\alpha)]}$     &  40.77   &  37.79   &  38.14   &  38.24   &  38.01   &  38.41   &  38.38   &  38.38   &  38.28  \\
$\log{[L(8.0)]}$         &  42.61   &  39.08   &  39.64   &  39.46   &  38.50   &  39.65   &  39.64   &  39.64   &  39.64  \\
$\log{[L(24)]}$          &  41.99   &  38.73   &  39.73   &  39.00   &  38.67   &  39.35   &  39.37   &  39.37   &  39.42  \\
\enddata
\tablecomments{Table 2 is available in its entirety via the link to the machine-readable version above. Units of right ascension are hours, minutes, and seconds, and units of declination are degrees, arcminutes, and arcseconds. The radius of each region is given in arcsec. All luminosities refer to observed values (no internal or Galactic extinction corrections have been applied) and the units are erg~s$^{-1}$.}
\end{deluxetable*}

%% file: tab3_2c.tex
\placetable{IRintegr_corrs}
\begin{deluxetable*}{llrrrrrr}
\scriptsize
\tablecaption{\label{IRintegr_corrs}Summary of correlations between mid-infrared monochromatic luminosities and colors and the integrated IR luminosities.}
\tablehead{\colhead{Estimator} & BKG & \colhead{A} & \colhead{$B(8.0)$} & \colhead{$B(24)$} & \colhead{$B(70)$} & \colhead{$B(160)$} & \colhead{$\sigma$}\\}
\startdata
$L(8-1000)$         & GLOBAL\footnote{This correlation is given in linear scale, not logarithmic.} & \multicolumn{1}{c}{\nodata}     & \multicolumn{1}{c}{\nodata}  & $+$1.439$\pm$0.415 & $+$0.814$\pm$0.099 & $+$1.1229$\pm$0.039 &  1\% \\
$\log{[L(8-1000)]}$ & GLOBAL & $+$0.63$\pm$0.11 & $+$1.020$\pm$0.015 & \multicolumn{1}{c}{\nodata}  & \multicolumn{1}{c}{\nodata}  &  \multicolumn{1}{c}{\nodata} &  8\% \\
                    & GLOBAL & $+$1.25$\pm$0.06 &  \multicolumn{1}{c}{\nodata} & $+$0.997$\pm$0.010 & \multicolumn{1}{c}{\nodata}  &  \multicolumn{1}{c}{\nodata} & 12\% \\
                    & GLOBAL & $+$1.71$\pm$0.46 &  \multicolumn{1}{c}{\nodata} & \multicolumn{1}{c}{\nodata}  & $+$0.839$\pm$0.065 &  \multicolumn{1}{c}{\nodata} &  9\% \\
                    & GLOBAL & $-$0.27$\pm$0.36 &  \multicolumn{1}{c}{\nodata} & \multicolumn{1}{c}{\nodata}  & \multicolumn{1}{c}{\nodata}  & $+$1.064$\pm$0.048 &  5\% \\
                    & GLOBAL & $+$0.94$\pm$0.05 & $+$0.644$\pm$0.102 & $+$0.356$\pm$0.102 & \multicolumn{1}{c}{\nodata}  &  \multicolumn{1}{c}{\nodata} &  6\% \\
$L(8-1000)$         & LOCAL$^a$  & \multicolumn{1}{c}{\nodata}     & \multicolumn{1}{c}{\nodata}  & $+$0.202$\pm$0.510 & $+$0.802$\pm$0.160 & $+$1.303$\pm$0.061 &  4\% \\
$\log{[L(8-1000)]}$ & LOCAL  & $+$0.95$\pm$0.13 & $+$0.980$\pm$0.019 & \multicolumn{1}{c}{\nodata}  & \multicolumn{1}{c}{\nodata}  &  \multicolumn{1}{c}{\nodata} &  7\% \\
                    & LOCAL  & $+$1.88$\pm$0.27 &  \multicolumn{1}{c}{\nodata} & $+$0.897$\pm$0.041 & \multicolumn{1}{c}{\nodata}  &  \multicolumn{1}{c}{\nodata} & 13\% \\
                    & LOCAL  & $+$1.78$\pm$0.27 &  \multicolumn{1}{c}{\nodata} & \multicolumn{1}{c}{\nodata}  & $+$0.824$\pm$0.038 &  \multicolumn{1}{c}{\nodata} & 11\% \\
                    & LOCAL  & $-$0.14$\pm$0.30 &  \multicolumn{1}{c}{\nodata} & \multicolumn{1}{c}{\nodata}  & \multicolumn{1}{c}{\nodata}  & $+$1.050$\pm$0.041 &  6\% \\
                    & LOCAL  & $+$0.93$\pm$0.06 & $+$0.695$\pm$0.155 & $+$0.305$\pm$0.155 & \multicolumn{1}{c}{\nodata}  &  \multicolumn{1}{c}{\nodata} & 11\% \\
$L(3-1100)$         & GLOBAL$^a$ & \multicolumn{1}{c}{\nodata}     & \multicolumn{1}{c}{\nodata}  & $+$1.832$\pm$0.420 & $+$0.779$\pm$0.094 & $+$1.176$\pm$0.031 &  1\% \\
$L(3-1100)$         & LOCAL$^a$  & \multicolumn{1}{c}{\nodata}     & \multicolumn{1}{c}{\nodata}  & $+$0.294$\pm$0.551 & $+$0.766$\pm$0.190 & $+$1.392$\pm$0.063 &  4\% \\
\enddata
\tablecomments{The Table shows the coefficients of the correlations 
(in logarithmic scale except for the correlations involving the 3 MIPS
wavelengths) between the MIR and FIR monochromatic luminosities and
the total IR luminosity integrated between 8 and 1000~\mic\,
[$L(8-1000)$] and 3 and 1100~\mic\, [$L(3-1100)$]. The equation for
each correlation is: $\log{[L(8-1000)]}=A+\sum_i
B(\lambda_i)\times\log{L(\lambda_i)}$ [except for the correlation
between the 3 MIPS luminosities and the integrated IR emission, where
the correlation is $L(8-1000)=\sum_i B(\lambda_i)\times
L(\lambda_i)$]. The sum extends over the wavelenghts of the last IRAC
channel and the 3 MIPS bands (8.0, 24, 70, and 160~\mic). In the last
column, $\sigma$ represents the scatter of the data around the given
relation. All luminosities are given in erg~s$^{-1}$.}
\end{deluxetable*}

%% file: tab4_2c.tex
\placetable{sfr_corrs}
\begin{deluxetable*}{lrrccrrc}
\tablecaption{\label{sfr_corrs}Summary of correlations between different SFR estimators.}
\tablehead{Estimator & \multicolumn{3}{c}{$\log{[L^\mathrm{corr}(\mathrm{H}\alpha)]}$} & & \multicolumn{3}{c}{$\log{[L^\mathrm{abs}(\mathrm{H}\alpha)]}$}\\
  \cline{2-4} \cline{6-8}\\
 & \multicolumn{1}{c}{a} & \multicolumn{1}{c}{b} & \multicolumn{1}{c}{$\sigma$} & & \multicolumn{1}{c}{a} & \multicolumn{1}{c}{b} & \multicolumn{1}{c}{$\sigma$}}
\startdata
$\log{[L(8-1000)]}$ &   2.3$\pm$1.2 & 0.992$\pm$0.031 & 30\% & & 8.9$\pm$1.0 & 0.826$\pm$0.025 & 30\% \\
$\log{[L(24)]}$     & -10.7$\pm$1.4 & 1.302$\pm$0.036 & 40\% & & 1.2$\pm$1.3 & 1.002$\pm$0.034 & 30\% \\
$\log{[L(8)]}$      &  -4.5$\pm$1.6 & 1.147$\pm$0.042 & 60\% & & 5.2$\pm$1.2 & 0.902$\pm$0.031 & 60\% \\
\enddata
\tablecomments{The Table shows the coefficients of the correlations between the luminosity given in 
the first column and $\log{L^\mathrm{corr}(\mathrm{H}\alpha)}$ or
$\log{L^\mathrm{abs}(\mathrm{H}\alpha)}$.  The equation for the
correlation is:
$\log{[L(\lambda)]}=a+b\times\log{[L^?(\mathrm{H}\alpha)]}$. In the
fourth and seventh columns of the Table, $\sigma$ is the scatter
around the linear relation. All luminosities are given in
erg~s$^{-1}$.}
\end{deluxetable*}